\documentclass[12pt]{article}

\usepackage{epsf}
\usepackage{amssymb}
\usepackage{a4}

\newcommand{\extraspace}{\addtolength{\abovedisplayskip}{2mm}
                        \addtolength{\belowdisplayskip}{2mm}
                        \addtolength{\abovedisplayshortskip}{2mm}
                        \addtolength{\belowdisplayshortskip}{2mm}}
\newcommand{\be}{\begin{equation}\extraspace}

\newcommand{\ee}{\end{equation}}

\newcommand{\bea}{\begin{eqnarray}\extraspace}
\newcommand{\beastar}{\begin{eqnarray*}\extraspace}
\newcommand{\eea}{\end{eqnarray}}
\newcommand{\eeastar}{\end{eqnarray*}}

\newcommand{\nonu}{\nonumber \\[2mm]}

\newcommand{\del}{\partial}

\newcommand{\bra}{\langle}
\newcommand{\ket}{\rangle}
\newcommand{\half}{{\textstyle \frac{1}{2}}}

\newcommand{\tablo}[1]{{\left\{#1\right\}}}
\newcommand{\jip}[2]{j^{#2}_{\tablo{#1}}}
\newcommand{\pip}[3]{p^{#2}_{\tablo{#1},#3}}
\newcommand{\Jjack}[3]{J_{\tablo{#1}}^{#2}\left(#3\right)}
\newcommand{\JEjack}[2]{J_{\tablo{#1}}^{#2}}
\newcommand{\Pjack}[3]{P_{\tablo{#1}}^{#2}\left(#3\right)}
\newcommand{\PEjack}[2]{P_{\tablo{#1}}^{#2} }
\newcommand{\Ojack}[3]{O^p_{\tablo{#1},#3}\left(#2\right)}
\newcommand{\Mmode}[2]{ M_{#1}(#2) }

\newcommand{\REpol}[1]{\mathcal{J}_{\tablo{#1}}}

\newcommand{\eps}{\epsilon}

\newcommand{\gt}{\tilde{g}}
\newcommand{\vac}{| 0 \rangle}

\begin{document}

\begin{center}

{\bf\LARGE Form Factors for Quasi-particles}

\vskip 3mm

{\bf\LARGE in $c=1$ Conformal Field Theory}

\vskip 8mm

{\Large R.A.J. van Elburg and K. Schoutens}

\vskip 5mm

{\sl
Van der Waals-Zeeman Institute and Institute for Theoretical Physics \\
University of Amsterdam, Valckenierstraat 65 \\
1018 XE Amsterdam, The Netherlands}

\end{center}

The non-Fermi liquid physics at the edge of fractional
quantum Hall systems is described by specific
chiral Conformal Field Theories with central charge
$c=1$. The charged quasi-particles
in these theories have fractional charge and obey a form of
fractional statistics. In this paper we study form factors,
which are matrix elements of physical (conformal) operators,
evaluated in a quasi-particle basis that is organized according
to the rules of fractional exclusion statistics. Using the
systematics of Jack polynomials, we derive selection rules
for a special class of form factors. We argue that finite temperature
Green's functions can be evaluated via systematic form factor
expansions, using form factors such as those computed in this paper and
thermodynamic distribution functions for fractional exclusion
statistics. We present a specific case study where we demonstrate
that the form factor expansion shows a rapid convergence.


\vfill

\noindent ITFA-2000-06


\noindent July 2000

%

\newpage

\section{Introduction and summary}
\setcounter{equation}{0}
\setcounter{figure}{0}

Shortly after the first observation of the fractional
quantum Hall (fqH) effect \cite{fqH}, R.B.~Laughlin proposed
that this phenomenon has its origin in a new state of matter,
which is formed by electrons but nevertheless admits excitations
of fractional charge \cite{Lau}.
The experimental evidence for the existence of fractionally
charged excitations includes the results of shot-noise measurements
for charge transport through fqH samples \cite{shot}.

The unusual properties of the fqH quasi-particles do not stop at
their fractional charge; in addition, the (bulk and edge)
quasi-particles exhibit various forms of fractional
quantum statistics \cite{Halp,ASW,Hal1,ICJ}.

In a previous paper \cite{vES} we initiated a program that aims at
computing various transport properties of fqH systems in a formalism
that makes direct reference to the (fractionally) charged quasi-particles.
In particular, we presented a quasi-particle basis
of the edge Conformal Field Theory (CFT) for the $\nu={1 \over p}$
principal Laughlin states. We demonstrated that the CFT quasi-particles
satisfy a form a fractional statistics that is closely related to
Haldane's notion of `fractional exclusion statistics' \cite{Hal1}.

The CFT for the $\nu=1/p$ fqH edges can be identified with 
a continuum ($N\to\infty$) limit of the Calogero-Sutherland (CS) 
model for quantum
mechanics with inverse square exchange. The natural quasi-particles
for the fqH system correspond to eigenstates of the CS hamiltonian.
In the work of many groups, the eigenstates of hamiltonians of the
CS-type have been understood in terms of Jack symmetric polynomials.
In particular, S.~Iso \cite{Is} presented two alternative
Jack polynomial bases for the continuum CS theory. In section 3.4
below, we discuss the precise correspondence between these CS bases 
and the fqH basis we presented in \cite{vES}. 

It has been recognized by many groups that quantum many
body systems with inverse square exchange come close to
being `ideal gases of fractional statistics particles'.
Supporting this claim are
the observations that (i) equilibrium thermodynamic
quantities can be evaluated with the help of 1-particle
distribution functions for fractional statistics and,
(ii) the zero-temperature correlation functions of simple
operators (such as electron and hole operators)
are mediated by intermediate states with a minimal
number of propagating quasi-particles [see section 4].

When turning to finite temperature, one quickly discovers
that the second `free gas' property (ii) no longer holds.
There are important many body effects which
can not be ignored when computing correlation functions
of physical operators at finite temperature.
This implies that great care needs to be taken when setting up
arguments that link the $T$-dependence of physical observables
($I$-$V$ and shot-noise characteristics in particular)
to the fractional statistics of the fundamental charge carriers
\cite{IMO}.

In this paper we turn to the problem of extending the
formalism based on quasi-particles with fractional
statistics to finite temperatures. We argue that specific
finite-$T$ correlation functions can be  written in a
so-called form factor expansion. Such expansions
start with a term that refers to a minimal number of
quasi-particles; to this leading term successive
corrections are added that refer to more and more
quasi-particles participating in physical process
described by the correlation function. To test the validity
of the proposed expansion, we explicitly
evaluate the form factor expansion for a specific finite-$T$
Green's function, collecting all terms that refer to
one or two quasi-particles. The numerical results
show a rather rapid convergence to the (known) exact
expression.

The form factor expansion that we propose is similar in spirit
to expressions that were proposed in \cite{LM}
(see also \cite{Sa}) in the
context of integrable quantum field theories whose
structure is set by a factorizable scattering matrix. Despite
this similarity, the two approaches are rather different:
in our approach we nowhere rely on scattering data
or on the associated form factor axioms, but instead perform
explicit computations in a CFT that is regularized by
the finite size $L$ of the spatial direction.

This paper is organized as follows.

In section 2 we review results obtained in our earlier
paper \cite{vES}. We present a basis for the edge theories
for the $\nu={1 \over p}$ fqH state,
employing edge electrons and edge quasi-holes as the
fundamental charged quasi-particles. We describe the
associated 1-particle thermodynamic distribution
functions and discuss the fundamental duality between
the two types of quasi-particles.

In section 3 we discuss the continuum CS models
and the associated Jack polynomial bases as first given
by Iso. We give an explicit 1-1 connection between
the states in the CS basis and the states in the fqH
basis.

In section 4 we turn to form factors. In 4.1 we
discuss the ones relevant for $T=0$, while in 4.2
we present simple examples of form factors that contribute
at non-zero $T$. The expressions that we obtain
make clear that the simple picture of `an ideal gas'
breaks down at finite temperature. A rather general
set of selection rules is presented in 4.3, while in section 4.4 we
briefly discuss the extent to which our explicit results
can be understood from an axiomatic approach based
on a factorized S-matrix.

In section 5 we compute 1 and 2-particle diagonal
form factors for the `edge electron counting operator'.
These form factors are then used to evaluate the
leading terms in a form factor expansion for the finite-$T$
Green's function that determines the $I$-$V$
characteristics of edge tunneling processes.

In section 6 we present our conclusions. In the
appendices we specify algebraic properties of the charged 
edge operators for $\nu={1 \over 2}$, we summarize some 
relevant results on Jack polynomials, and we provide a 
proof of the form factor selection rules presented in 
section 4.3.

\section{Quasi-particle basis of fqH edge theories}
\setcounter{equation}{0}
\setcounter{figure}{0}

\subsection{fqH edge theories as CFT}

It is well known that many (though not all!) aspects of
the low energy dynamics of fqH systems can be captured by
an effective edge theory. These edge theories are
so-called chiral Luttinger Liquids or chiral CFTs. For
the specific case of a principal Laughlin state at filling
$\nu=1/p$, the CFT describing a single edge (in isolation)
is a specific $c=1$ CFT.

The standard interpretation of this effective theory is in
terms of a chiral boson, which is identified with a quantized
density wave (magnetoplasmon) along the edge of the fqH
sample. By exploiting the relation between the chiral anomaly
and the quantized Hall conductance, one finds that the
chiral boson field is compactified on a circle of radius $R^2=p$
\cite{We1}. This construction directly leads to a space of states
(the so-called chiral Hilbert space) with partition function
\be
Z^{1/p}(q) = \sum_{Q=-\infty}^{\infty} {q^{Q^2 \over 2p}
                 \over (q)_{\infty}} \ ,
\label{Zp}
\ee
with $(q)_\infty = \prod_{l=1}^{\infty}(1-q^l)$ and
$q=e^{-\beta {2\pi \over L}{1 \over \rho_0}}$. [The 1-particle
energies are spaced by $l {2\pi \over L}{1 \over \rho_0}$
with $l$ integer and $\rho_0$ the density of states per unit length,
$\rho_0=(\hbar v_F)^{-1}$.]

In eq.~(\ref{Zp}), the parameter $Q$ is the electric $U(1)$ charge
in units of ${e \over p}$. The Hilbert space is obtained as a collection
of charge sectors $Q$. Within each sector, there is a leading
state of minimal energy ${Q^2 \over 2p}$; all other states in that
sector are reached via the application of (neutral) Fourier modes of
the density operator. Together, these modes form a $U(1)$ Kac-Moody
algebra, and the factor ${1 \over (q)_\infty}$ is the well-known character
of a highest weight module of this affine algebra.

In our earlier paper \cite{vES}, we proposed that the CFT for
the $\nu={1 \over p}$ fqH edge can be interpreted in terms of 
a set of  fundamental
charged quasi-particles. We have worked out a formulation in terms of
edge electrons (of charge $-e$) and edge quasi-holes (of charge
$+{e \over p}$). Our main motivation has been that, using this novel
formulation, one learns how to understand some of the unusual and
spectacular phenomenology of the fqH systems as manifestations of
unusual properties (fractional charge and fractional statistics in
particular) of their fundamental quasi-particles.

\subsection{The fqH quasi-particle basis}

In \cite{vES}, we demonstrated how the collection of states
(\ref{Zp}) can be understood as a collection of multi-particle
states, the fundamental (quasi-)particles being the edge
electron and the edge quasi-hole, of charge $Q=-p$ and $Q=+1$,
respectively.

The edge electron and quasi-holes are described by the
conformal, primary fields
\be
J^{(-p)}(z) = \sum_t J^{(-p)}_{-t} z^{t-{p \over 2}} \ ,
\qquad
\phi^+(z) = \sum_s \phi^+_{-s}  z^{s-{1 \over 2p}}  \ .
\ee
Clearly, one can employ the Fourier modes $ J^{(-p)}_{-t}$
and $\phi^+_{-s}$ as `creation operators' for the corresponding
quasi-particles. [The reason why we put quotation marks
here will soon become clear.] In \cite{vES} we identified a
collection of multi-$J$, multi-$\phi$ states which together
form a basis for the chiral Hilbert space. It is given by the states
(from now on we omit the charge superindex on the operators $J$,
$\phi$)
\bea
&&
 J_{-(2M-1){p \over 2}+Q -m_M} \ldots
 J_{-{p \over 2}+Q -m_1}
 \phi_{-(2N-1){1 \over 2p}-{Q \over p}-n_N} \ldots
 \phi_{-{1 \over 2p}-{Q \over p}-n_1} | \, Q \, \rangle
\nonu
&& \quad {\rm with}\ \quad
   m_M \geq m_{M-1} \geq \ldots \geq m_1 \geq 0 ,\quad
   n_N \geq n_{N-1} \geq \ldots \geq n_1 \geq 0  \ ,
\nonu
&& \qquad \qquad  n_1 >0 \quad {\rm if}\ \ Q < 0 \ ,
\label{listofstates}
\eea
where $|Q\rangle$ denotes the lowest energy state of charge
$Q {e \over p}$, with $Q$ taking the values $-(p-1)$,
$-(p-2)$, $\ldots$, $-1$, $0$.

The associated character identity is
\be
  Z^{1/p}(q)= \sum_{Q=-(p-1)}^{0} \, q^{Q^2\over 2p} \,
           Z^{\rm qh}_Q(q) \, Z^{\rm e}_Q(q) \ ,
\label{Zid}
\ee
where the factor $q^{Q^2\over 2p}$ takes into account the
energy of the initial states and we denoted by $Z^{\rm qh}_Q$,
$Z^{\rm e}_Q$ the partition functions for quasi-holes and
electrons in
the sector with vacuum charge $Q$. They are naturally written as
\be
Z^{\rm qh}_Q = \sum_{N=0}^\infty
{q^{ {1 \over 2p}(N^2+2QN)+(1-\delta_{Q,0})N } \over (q)_N} \ ,
\qquad
Z^{\rm e}_Q = \sum_{M=0}^\infty
{q^{ {p \over 2} M^2 -QM }  \over (q)_M} \ ,
\ee
with $(q)_L=\prod_{l=1}^L (1-q^l)$. The identity (\ref{Zid})
can be rigorously established by employing partition counting
theorems that are available in the mathematical literature
(see \cite{BM} for a discussion).

An important special case is $p=1$, where the quasi-particle
basis (\ref{listofstates})
is the standard multi-particle basis in a theory of free,
charged but spin-less fermions.

\subsection{Fractional statistics and duality}

\subsubsection{Fractional exclusion statistics}

In a 1991 paper \cite{Hal1},  F.D.M.~Haldane
proposed the notion `fractional exclusion statistics',
as a tool for the analysis of strongly correlated
many-body systems. The central assumption that
is made concerns the way a many-body spectrum is
built by filling available one-particle states. In words, it
is assumed that the act of filling a one-particle state
effectively reduces the dimension of the space of remaining
one-particle states by an amount $g$. The choices $g=1$, $g=0$
correspond to fermions and bosons, respectively. The
thermodynamics for general `$g$-ons', and in particular the
appropriate generalization of the Fermi-Dirac distribution
function, have been obtained in \cite{IOW}. The so-called
IOW equations
\be
\bar{n}_g(\eps) = {1 \over [w(\eps)+g]} \ , \quad 
{\rm with} \quad
[w(\eps)]^g[1+w(\eps)]^{1-g} = e^{\beta(\eps-\mu)}
\label{wueq}
\ee
provide an implicit expression for the 1-particle distribution
function $\bar{n}_g(\eps)$ for $g$-ons at temperature $T$ and
chemical potential $\mu$. The solutions $\bar{n}_g(\epsilon)$
have limiting value $\bar{n}^{\rm max}_g={1 \over g}$ for
$\epsilon \to -\infty$.

\subsubsection{Spectral shift statistics and the fqH effect}

In \cite{vES} we analyzed the exclusion statistics behind the
states (\ref{listofstates}) that form a basis for the chiral
Hilbert space for a $\nu={1 \over p}$ fqH edge. This analysis
employed a technique, first proposed in \cite{Sc}, based on
recursion relations satisfied by truncated chiral
partition functions. The remarkable conclusions are that
\begin{itemize}
\item
the `microscopic' state-filling rules differ from those formulated
by Haldane, but
\item
the state counting, and thereby the 1-particle thermodynamic
distribution functions, agree with those
associated to fractional exclusion statistics, {\it i.e.}\ the
distribution functions are solutions of the IOW equations.
\end{itemize}
The precise statement is that the edge electrons are described by
the IOW distribution with $g=p$, while the edge quasi-hole states
are thermally occupied according to the distribution with
$\tilde{g}=1/p$. It is important to remark that there is no mutual
statistics between the two types of excitations.

For later reference, we list the explicit expressions for the distribution
functions for the case $p=2$
\be
\bar{n}_2(\eps) = \half \left(
1 - {1 \over \sqrt{ 1+4\, e^{-\beta(\eps-\mu)}}} \right) \ ,
\quad
\bar{n}_{\half}(\eps) = {2  \over  \sqrt{1+4e^{2\beta(\eps-\mu)}}} \ .
\ee

\subsubsection{Duality}

The distribution functions for fractional exclusion
statistics with parameters $g$ and $\tilde{g}=1/g$
satisfy the following duality relation \cite{dual}
\be
g \, \bar{n}_g(\eps) = 1 - \gt \, \bar{n}_{\gt}(-\gt \eps) \ .
\label{dual}
\ee
The interpretation of this result is that the $\gt$ quanta
with positive energy act as holes in the ground state
distribution of negative energy $g$-quanta.

Translating back to the $\nu={1 \over p}$ fqH edge,
we observe a fundamental duality between edge electrons and
edge quasi-holes, in agreement with the physical
interpretation of strong and weak backscattering limits of
edge-to-edge tunneling processes \cite{FLS}. Under the duality,
the removal of a single edge electron corresponds to the
creation of a total of $p$ quasi-holes.

This duality further implies
that, when setting up a quasi-particle description for
fqH edges, we can opt for (i) either quasi-holes
\`or edge electrons, with energies over the full range
$-\infty < \eps < \infty$ (`particle picture'),
or (ii) a combination of both
types of quasi-particles, each having positive energies
only (`excitation picture'). The option (ii) is the one realized
in the fqH basis of section 2.2. In section 3 we shall discuss
the CS bases proposed by Iso, which in a sense uses the
option (i).

\subsection{Equilibrium quantities}

\subsubsection{Specific heat}

The specific heat of a conformal field theory is well-known to
be proportional to the central charge $c_{CFT}$
\be
{C(T) \over L}
= \gamma \rho_0 k_B^2 T \ , \quad
\gamma = {\pi \over 6} \, c_{CFT} \ ,
\ee
where $\rho_0=(\hbar v_F)^{-1}$ is the density of
states per unit length. In \cite{vES} we demonstrated that
the fqH quasi-particle basis specified in (\ref{listofstates})
leads to (with $g=p$, $\tilde{g}={1 \over p}$)
\be
\gamma  =  \gamma_{g,+} + \gamma_{\gt,+}
\ee
with
\be
\gamma_{g,+} = \del_\beta \int_0^\infty
  d\eps \, \eps \, \bar{n}_g(\eps) \ ,
\quad
\gamma_{\gt,+} = \del_\beta \int_0^\infty
  d\eps \, \eps \, \bar{n}_{\gt}(\eps) \ .
\ee

One finds that, while the individual contributions
$\gamma_{g,+}$, $\gamma_{\tilde{g},+}$
depend on  $g$, their sum is equal to
${\pi \over 6}$ for all $g$, in agreement with $c_{\rm CFT}=1$.
For $g=2$, $\tilde{g}=\half$ one has
\be
\gamma_{2,+} = {\pi \over 6} \, {2 \over 5} ,\quad
\gamma_{\half,+} = {\pi \over 6} \, {3 \over 5}  \ .
\ee

\subsubsection{Hall conductance}

By a simple back-of-the-envelope argument, the
Hall conductance is related to the edge capacitance,
{\it i.e.} to the charge $\Delta Q$ that is accumulated
on a given edge in response to an applied voltage $V$.
One quickly derives the following expression for the
Hall conductance
\be
G/[{e^2 \over h}] = {1 \over eV} \left[
- \int_0^\infty d\eps  \, \bar{n}_p(\eps + e V)
+ {1\over p} \int_0^\infty d\eps \,
\bar{n}_{{1 \over p}}(\eps-{e \over p}V) \right] \ .
\label{gformula}
\ee
Using the properties of the distribution functions,
one shows that this expression is independent of
the temperature, and gives $G={1 \over p}{e^2 \over h}$
for the $\nu={1 \over p}$ edge. For $T=0$
eq.~(\ref{gformula}) reduces to
\be
G = \bar{n}^{\rm max}_g \, {q^2 \over h} \ ,
\ee
with $q$ the charge and $g$ the statistics parameter of
the quasi-particles that are pulled into the edge by the applied
voltage. Depending on the sign of $V$ these are the
edge electrons ($q=-e$, $\bar{n}^{\rm max}={1 \over p}$) or
the quasi-holes ($q={e \over p}$, $\bar{n}^{\rm max}=p$); both
give the canonical value of the Hall conductance.

\section{CS models and fractional statistics}
\setcounter{equation}{0}
\setcounter{figure}{0}

\subsection{Inverse square exchange in the CFT setting}

While the states specified in (\ref{listofstates})
form a complete basis of the chiral Hilbert space, they
are not mutually orthogonal, and as such they do not form
a proper starting point for further analysis. In principle
one may go through a orthogonalization procedure to
arrive at a proper canonical quasi-particle basis. We shall
here reach this goal in a more efficient way, by exploiting
the close connection with so-called Calogero-Sutherland
(CS)  models of many-body quantum mechanics.

The CS model describes the (non-relativistic) quantum mechanics
of $N$ particles on a circle, with 2-body interaction that is
proportional to the inverse square of the chord distance between
the particles. In \cite{Is}, S.~Iso demonstrated that the limit
$N\to \infty$ of a CS model with interaction strength
$p(p-1)$ can be identified with the $c=1$
CFT of a chiral boson compactified on a circle with radius $R^2=p$.
Iso also specified a collective hamiltonian $H_{CS}$, acting
in the CFT Hilbert space,  whose eigenstates precisely
correspond to the multi-particle states of the underlying CS model.

It turns out that the eigenstates of $H_{CS}$ are in 1-1
correspondence with the states specified in eq.~({\ref{listofstates}):
by adding subleading terms to the expressions in (\ref{listofstates}),
one arrives at a (orthogonal and complete) set of eigenstates of
$H_{CS}$. Comparing with Iso's formulation, one finds
that the `superfermions' of \cite{Is} correspond to what we call
edge electrons and the `anyons' of \cite{Is} are the edge
quasi-holes of the fqH system. Nevertheless, the `CS basis'
specified by Iso differs from the `fqH basis' of this paper through
the way in which the quasi-particle content of a given state is
specified. In subsection 3.4 below we shall spell out the precise
correspondence between the two formulations.

The spectrum of the CS models has been analyzed with the
help of so-called Jack symmetric polynomials, which
provide explicit wave functions and eigenstates for
the CS hamiltonian. With the help of  `Jack technology',
important conjectures \cite{Hal3}
about zero-temperature correlation functions
of models with inverse square exchange have been proven
\cite{Ha,LPS}. In the present paper, where we are interested in
finite-temperature correlation functions, we shall apply the
`Jack technology' to obtain a set of selection rules on
form factors that are relevant for computations at finite
temperature. We shall complement these
considerations with explicit computations of form factors involving
states with up to two quasi-particles.

\subsection{The hamiltonian $H_{CS}$ and the fqH basis}

To specify the operator $H_{CS}$, we employ a free chiral
scalar field $\varphi(z)$. In terms of this scalar field,
the charged fields $J$ and $\phi$ take the form of so-called
vertex operators,
\be
J(z)=e^{-i\sqrt{p}\varphi}(z) \ , \qquad
\phi(z)=e^{i{1 \over \sqrt{p}}\varphi}(z) \ .
\ee
The operator $Q = \oint (i\sqrt{p}\del\varphi)$ measures the 
electric charge in units ${e \over p}$.
Following \cite{Is}, we define
\be
   H_{CS} =
      {p-1 \over p} \sum_{l=0}^\infty (l+1) (i\sqrt{p}
                \del \varphi)_{-l-1} (i\sqrt{p}\del \varphi)_{l+1}
    + {1 \over 3p} \left[(i\sqrt{p} \del \varphi)^3\right]_0 \ ,
\label{Hcs}
\ee
where $\del \varphi(z) = \sum_l (\del \varphi)_l z^{-l-1}$
and where the second term on the r.h.s. denotes the zero-mode
of the normal ordered product of three factors
$(i \sqrt{p} \del \varphi)(z)$. As a first result, one
finds the following action of $H_{CS}$ on states
containing a single quasi-particle of charge $Q=+1$
or $Q=-p$
\bea
&&H_{CS}\, \phi_{-{1 \over 2p}-n} \vac =
  h_{\phi}(n) \, \phi_{-{1 \over 2p}-n} \vac
  \ , \qquad
h_{\phi}(n) = \left[ {1 \over 3p} + pn (n+{1 \over p})\right] 
\nonu
&&H_{CS}\, J_{-{p \over 2}-m} \vac =
  h_J(m) J_{-{p \over 2}-m} \vac
  \ ,  \qquad
h_J(m) = \left[ - {p^2 \over 3} - m (p+m) \right] \ .
\nonu
&&
\eea
We would like to stress that the fact that both $J_s$
and $\phi_t$ diagonalize $H_{CS}$ is quite non-trivial.
If one evaluates $H_{CS}$ on any vertex operator $\phi^{(Q)}$
(of charge $Q {e \over p}$), one typically runs into the field
product $(T\phi^{(Q)})(z)$, where $T(z)= - \half (\del \varphi)^2(z)$
is the stress-energy of the scalar field $\varphi$. Only for
$Q=1$ and $Q=-p$ do such terms cancel and do we find that the
quasi-particle states are eigenstates of $H_{CS}$.

We can now continue and construct eigenstates of $H_{CS}$
which contain several $\phi$ or $J$-quanta. What one
then finds is that the simple product states specified in 
(\ref{listofstates}) are not $H_{CS}$ eigenstates, but that they
rather act as head states that need to be supplemented by
a tail of subleading terms. For the $H_{CS}$ eigenstate headed by
the multi-particle state (\ref{listofstates}) (with unit coefficient),
we shall use the notation
\be 
| \{ m_j ; n_i\} \rangle^Q
\label{fqHstates}
\ee
so that
\bea
\lefteqn{
H_{CS} | \{ m_j ; n_i\} \rangle^Q =}
\nonu
&& \left[ {Q^3 \over 3p} +
 \sum_{j=1}^M h_J((j-1)p-Q+m_j)
  + \sum_{i=1}^N h_{\phi}({1 \over p}(Q+i-1)+n_i) \right]
 | \{ m_j ; n_i\} \rangle^Q .
\nonu &&
\label{Hcseigenvalue}
\eea
The states (\ref{fqHstates}), with the $m_j$, $n_i$ and $Q$
as specified in and below (\ref{listofstates}), form a complete
and orthogonal basis for the chiral Hilbert space.

For the sake of illustration, we give explicit results
for a few simple states of the fqH basis.
The 1-particle states over the $Q=0$ vacuum are given by
\be
|\{ m_1 \} \ket^0 = J_{-{p \over 2}-m_1}|0 \ket \ ,
\quad
|\{ n_1 \} \ket^0 = \phi_{-{1 \over 2p}-n_1}|0 \ket \ .
\ee
The norms of these states can explicitly be evaluated
by exploiting the (generalized) commutation relations satisfied
by the modes $J_s$ and $\phi_t$. One finds
\be
{}^0\bra \{m_1\}|\{m_1\}\ket^0 = C^{(-p)}_{m_1} \ ,
\quad
{}^0\bra \{n_1\}|\{n_1\}\ket^0 = C^{( -{1\over p})}_{n_1} \ ,
\ee
where the $C_k^{(\alpha)}$ are the  expansion coefficients of
$(1-x)^{\alpha}=\sum_{k\geq 0} C_k^{(\alpha)} x^k$.

In our discussion below we shall often restrict ourselves
to the vacuum sector $Q=0$, and omit the explicit sector
label $Q$ on the fqH states (\ref{fqHstates}).

For later use, we present the explicit form of the
fqH states $| \{m_2, m_1 \} \rangle$ and $| \{m_1;n_1\} \rangle$
at $p=2$
\bea
&& |\{ m_2,m_1\}\ket
   = |m_2,m_1\ket + {2\over m_2-m_1+3}\sum_{l>0} | m_2+l,m_1-l\ket \ ,
\label{jjstate}
\\[2mm]
&& |\{ m_1;n_1\}\ket
   = |m_1;n_1\ket
    -{1 \over (m_1+2 n_1 +1)} \sum_{l > 0}|m_1+l;n_1-l\ket \ ,
\label{jphistate}
\eea
with
\be
 |m_2,m_1\ket=J_{-3-m_2}J_{-1-m_1}|0\ket \ ,
 \qquad
 |m_1;n_1\ket=J_{-1-m_1}\phi_{-{1\over 4}-n_1}|0\ket \ .
\ee
By explicit evaluation, we obtain the following norms for these
states (again at $p=2$)
\bea
&& N_{\{m_2,m_1\}} =
\bra \{m_2,m_1\}|\{m_2,m_1\}\ket
= {m_2-m_1+1 \over m_2-m_1+3} (m_2+3)(m_1+1) \ ,
\nonu
&& N_{\{m_1;n_1\}} =
\bra \{m_1;n_1\}|\{ m_1;n_1\}\ket=
   {m_1+2n_1+2\over m_1+2n_1+1}
   (m_1+1) C_{n_1}^{(-{1\over 2})} \ .
\label{2partnorms}
\eea

\subsection{Jack polynomials and the CS bases}

In the previous section, we specified a complete set of
eigenstates of the hamiltonian $H_{CS}$ in terms of charged
quasi-particles $J$ and $\phi$. It is clear that these
same eigenstates can be obtained by applying an appropriate
polynomial in the modes $a_n=(\del\varphi)_n$ of the
auxiliary scalar field to a vacuum state $|q\rangle$.
It turns out that the polynomials that are needed are
so-called Jack polynomials. In appendix B we briefly
discuss some of their relevant properties.

Following Iso \cite{Is}, we may specify a basis of CS
eigenstates as follows
\be
 | \{ \mu \}_J, q \ket  =
 J_{\{\mu^\prime\}}^{{1 \over p}}(\{ \sqrt{p} a _{-n} \}) | q \ket
 = J_{\{\mu^\prime\}}^{{1 \over p}} | q \ket \ ,
\label{jackstate}
\ee
with the $U(1)$ charge $q$ running over all integers, and $\{\mu\}$
running over all Young tableaus. The norms of these states are given by
\be
\bra \{ \mu \}_J , q | \{ \mu \}_J , q \ket
= j_{\mu'}^{1 \over p} \ .
\label{jnorm}
\ee
An alternative `dual' basis, consists of the states
\be
 | \{ \nu \}_\phi , q \ket  =
 J_{\{\nu^\prime\}}^p ( \{ {a_{-n} \over \sqrt{p}} \}) \, | q \ket
 = J_{\{\nu^\prime\}}^p \, | q \ket \ ,
\ee
with $q$ integer and $\{\nu\}$ running over all Young tableaus,
with norms given by
\be
\bra \{ \nu \}_\phi , q | \{ \nu \}_\phi , q \ket
= j_{\nu'}^p \ .
\label{phinorm}
\ee

\subsection{Correspondence between fqH and CS bases}

Knowing that both the fqH quasi-particle basis and
the CS Jack polynomial basis (\ref{jackstate}) are complete bases 
of orthogonal eigenstates of the operator $H_{CS}$, one 
quickly concludes that there is a 1-1 identification between these 
two bases. In this section we explicitly describe this 1-1 
correspondence.

We start by observing that the Jack state
\begin{equation}
  | \tablo{\mu}_J,q \rangle=\JEjack{\mu'}{{1\over p}}  |q\rangle
\end{equation}
can be written as the sum of a leading state
\begin{equation}
  J_{-m_{\tilde{M}}+q+{p\over 2}}J_{-m_{\tilde{M}-1}+q+{3p\over
      2}} \dots J_{-m_1+q+{(2\tilde{M}-1)p\over 2} }|q+\tilde{M} p \rangle \ ,
\label{Jq-states}
\end{equation}
and a `tail' of sub-leading corrections, where `subleading' refers
to the triangular form of $H_{CS}$ on states of the form
(\ref{Jq-states}). Here we identified $m_j=\mu_{\tilde{M}+1-j}\geq 1$,
with $\tilde{M}=l(\tablo{\mu})$.

Similarly we identify
\begin{equation}
  |\tablo{\nu}_\phi,q\rangle=\JEjack{\nu'}{p}|q\rangle
\end{equation}
with the $H_{CS}$ eigenstate headed by
\begin{equation}
  \phi_{ -n_{\tilde{N}} -{q \over p} +{1\over 2p} } 
  \phi_{ -n_{\tilde{N}-1} -{q \over p} +{3\over 2p}}
  \dots \phi_{-n_1-{q \over p}+{2\tilde{N}-1\over 2p} }
  |q-\tilde{N}\rangle \ ,
\label{Jphi-states}
\end{equation}
with $n_i=\nu_{\tilde{N}+1-i}\geq 1$ and $\tilde{N}=l(\tablo{\nu})$.

Note that, as they stand, the expressions (\ref{Jq-states}) and
(\ref{Jphi-states}) are, in general, not of the form
(\ref{listofstates}) that defines a member of the fqH basis.

Using the above, we find the following identifications
for fqH states which contain only one type of mode operator
\begin{eqnarray}
&& | \{n_i\} \rangle^Q = | \tablo{\nu}_\phi, Q+N\rangle
\nonu
&& | \{ m_j\} \rangle^Q = | \tablo{\mu}_J, Q-pM \rangle \ ,
\end{eqnarray}
with $\nu_i=n_{N-i+1}\geq 1$ for $\tilde{N}-i \geq 0$ and
similarly  $\mu_j=m_{M-j+1}\geq 1$ for $\tilde{M}-j\geq 0$.
Note that only the $n_i$ that are non-zero become a
part of the tableau $\{\nu\}$; the $\phi$-modes with
$n_i=0$ change the charge of the vacuum without exciting
any of the $(\del \varphi)_n$ modes. [A similar remark applies
to the $J$-modes with $m_j=0$.]

The duality eq.~(\ref{jackdual}) on the Jack polynomials leads to the
following duality relation for the Jack operators,
\begin{equation}
  \JEjack{\lambda'}{p}=(-1)^{|\lambda|}{
    \jip{\lambda}{p}  }\JEjack{\lambda}{1\over p} \ .
\end{equation}
This relation enables us rewrite the states $ | \{n_i\}
\rangle^Q$ to either the $|\tablo{\nu'}_J,Q+N\rangle$ or the
$|\tablo{\nu}_\phi,Q+N\rangle$ form.
We can for example rewrite
\begin{eqnarray}
| \{n_i\} \rangle^Q &=& |\tablo{\nu}_\phi,Q+N\rangle\nonu
&=&{(-1)^{|\nu|} \jip{\nu'}{{p}}  } |\tablo{\nu'}_J,Q+N\rangle \ .
\label{QtoQN}
\end{eqnarray}
The last state is equivalent to the state which has
\begin{equation}
 J_{-\nu'_1+Q+N+{p\over 2}}J_{-\nu'_2+Q+N+{3p\over
         2}}...J_{-\nu'_{n_N}+Q+N+{(2n_N-1)p\over 2}}|Q+N+ n_N p \rangle
\label{dualJstate}
\end{equation}
as its leading state (assuming $n_1>0$).

One may explicitly check that the eigenvalues of both the Virasoro 
zero-mode $L_0$ and and the CS hamiltonian $H_{CS}$ are invariant
under the duality transformation that identifies the state
headed by (\ref{Jphi-states}) to the state headed by (\ref{dualJstate}).

If there are no $J$-operators present in the fqH head state we can 
use the duality to transform the $\phi$-operators in the head state into
$J$-operators and we achieve our goal of identitying the fqH state
with a member of the CS $J$-basis. If the fqH state has both $J$ and 
$\phi$-operators present, we can still map the $\phi$-operators to a 
dual set of $J$-operators. Starting from the state 
$| \{m_j\}, \{n_i\} \rangle^Q$
we see that, upon using the identity (\ref{QtoQN}), the $J$-modes
associated to the $m_j$ `see' their vacuum shifted from
$|Q\rangle$ to $|Q+N\rangle$. Since the vacuum charge leads
to a shift in the values of the mode-indices (see (\ref{Jq-states})),
this means that in the CS basis, the corresponding $J$-modes will
be labeled by $m_i+N$ instead of $m_i$. It is important to remark
that this shift does not affect the contribution of the 
$J$ modes to the eigenvalue of $H_{CS}$ on the state. This is 
because the eigenvalues of $H_{CS}$, as specified
in eq.~(\ref{Hcseigenvalue}), depend directly on the full 
mode-indices in the head state and not just on the labels
$m_j$, as can be seen by comparing eq.(\ref{Hcseigenvalue})
with (\ref{listofstates}).

We can now give the full mapping from a fqH basis state
to a state in the CS basis
\begin{eqnarray}
 | \{ m_j ; n_i\} \rangle^Q =
|\tablo{\sigma}_J,Q+{N}-p{M}\rangle \nonumber\\[4mm]
{\rm with}\;\; \tablo{\sigma}=(\tablo{m}+N^M)\cup \tablo{\nu'} \ ,
\label{fqHtoCS}
\end{eqnarray}
where the sum of the partitions is
$\tablo{\mu}+N^M=(\mu_1+N,\mu_2+N,..,\mu_M+N)$, and the cup product
$\tablo{\lambda}\cup\tablo{\rho}$ denotes the partition obtained from
sorting the parts $(\lambda_1,...,\lambda_S,\rho_1,...,\rho_R)$ in
descending order.

\setlength{\unitlength}{1 em}
\begin{figure}[t]
  \begin{picture}(15,17)(0,0)
\put(4,10){$\tablo{\mu}$}   
\multiput(0,7)(1,0){8}{\framebox(1,1){}}
\multiput(0,6)(1,0){8}{\framebox(1,1){}}
\multiput(0,5)(1,0){5}{\framebox(1,1){}}
\put(9,6){,}
\put(13,10){$\tablo{\nu}$}       
\multiput(10,7)(1,0){6}{\framebox(1,1){}}
\multiput(10,6)(1,0){4}{\framebox(1,1){}}
\multiput(10,5)(1,0){4}{\framebox(1,1){}}
\multiput(10,4)(1,0){4}{\framebox(1,1){}}
\put(17,7){$\rightarrow$}
\put(23.5,13){$N$}
\put(24,12.5){\vector(1,0){3}}
\put(24,12.5){\vector(-1,0){2}}
\put(19.5,9){$M$}
\put(21.5,9){\vector(0,1){3}}
\put(21.5,9){\vector(0,-1){2}}
\put(30,13){$\tablo{\mu}$}
\multiput(22,11)(1,0){13}{\framebox(1,1){}}
\multiput(22,10)(1,0){13}{\framebox(1,1){}}
\multiput(22,9)(1,0){10}{\framebox(1,1){}}
\multiput(22,8)(1,0){5}{\framebox(1,1){}}
\multiput(22,7)(1,0){5}{\framebox(1,1){}}
\multiput(22,11)(1,0){5}{\framebox(.75,.75){}}
\multiput(22,10)(1,0){5}{\framebox(.75,.75){}}
\multiput(22,9)(1,0){5}{\framebox(.75,.75){}}
\multiput(22,8)(1,0){5}{\framebox(.75,.75){}}
\multiput(22,7)(1,0){5}{\framebox(.75,.75){}}
\put(19.5,4){$\tablo{\nu'}$}
\multiput(22,6)(1,0){4}{\framebox(1,1){}}
\multiput(22,5)(1,0){4}{\framebox(1,1){}}
\multiput(22,4)(1,0){4}{\framebox(1,1){}}
\multiput(22,3)(1,0){4}{\framebox(1,1){}}
\multiput(22,2)(1,0){1}{\framebox(1,1){}}
\multiput(22,1)(1,0){1}{\framebox(1,1){}}

  \end{picture}
  \caption{Mapping from the state
    $|({8,8,5,0,0}),({6,4,4,4,0})\rangle^0$
    in the fqH basis to the state
    $|(13,13,10,5,5,4,4,4,4,1,1)_J,q=5-5p \rangle$
    in the CS basis.}
  \label{fig:fqhtocs}
\end{figure}
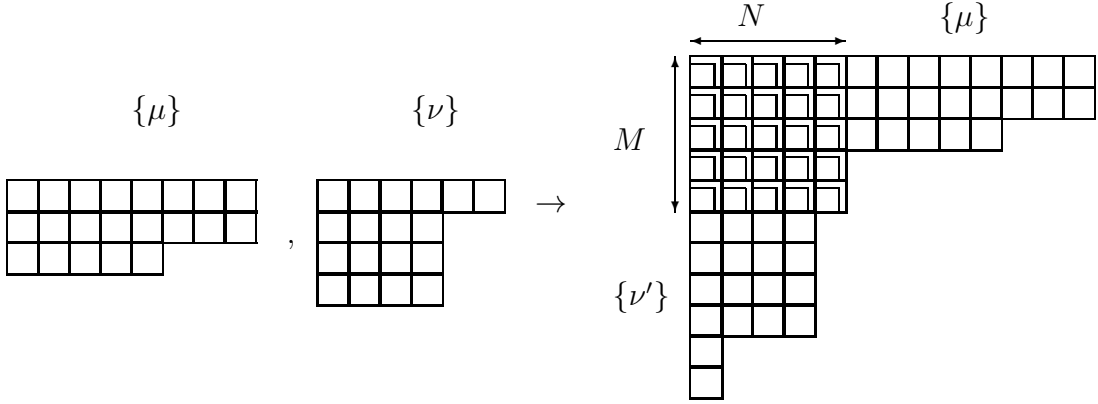

The mapping from the CS basis back to the fqH basis is slightly more
complicated. We start from a CS state $|\tablo{\sigma}_J,q\rangle$,
with $S=l(\tablo{\sigma})$,
to which we associate a multi-$J$ state as in (\ref{Jq-states}).
We note that the quantity $\sigma_j-q-pj$ decreases with increasing
label $j$ which allows us to fix $j$ such that
\begin{equation}
\sigma_j-q-pj \geq 0 \ , \qquad
\sigma_{j+1}-q-p(j+1) < 0 \ .
\label{j-cond}
\end{equation} 
The condition on $\sigma_j$ guarantees that the associated
$J$-mode is an allowed mode on a fqH vacuum $|Q\rangle$ with
$-(p-1)\leq Q\leq 0$. We now consider the state that remains
when all $J$-modes left from and at the position $j$ are removed.
With the use of duality this state can be rewritten,
\begin{equation}
 |\tablo{\nu'}_J,q+pj\rangle \propto |\tablo{\nu}_\phi,q+pj\rangle
\end{equation}
with
\begin{equation}
\nu'_i=\sigma_{i+j} \;\; {\rm for}\;\; i=1\dots S-j \ .
\label{nuprime}
\end{equation}
The $\phi$-operators in the leading state of
$|\tablo{\nu}_\phi,q+pj\rangle$ act on the vacuum with charge
$\tilde{Q}=q+pj-\sigma_{j+1}$ vacuum. The second condition in
(\ref{j-cond}) guarantees that $\tilde{Q}\geq -(p-1)$. If one
has $\tilde{Q} \leq 0$, one identifies $Q=\tilde{Q}$ as
the vacuum charge of the fqH basis state. If, however, $\tilde{Q}$
is larger than zero the state $|\tilde{Q}\rangle$ was created from
$|0\rangle$ using $\phi$-operators with the highest allowed mode
index. From this argument we obtain that the state
$|\tablo{\sigma}_J,q \rangle$ from the CS basis can be rewritten
as $|\{m_i;n_i\}\rangle^Q$ with the following rules for selecting
$Q$, $M$ and $N$ and the mode indices $\{m_i\}$ and $\{n_i\}$
\begin{equation}
  M=j \ , \quad N={\rm max}(q+pj,\sigma_{j+1}) \ , \quad Q=q+pM-N
\end{equation}
\begin{eqnarray}
m_i & =\sigma_{j+1-i}-N \quad & {\rm for} \  i=1\dots j
\nonu
n_i & = \nu_{\sigma_{j+1}+1-i} \quad & {\rm for} \ i=1\dots \sigma_{j+1}
\nonu
    & = 0 \qquad\qquad & {\rm for} \ i=\sigma_{j+1}+1 \ldots N \ ,
\end{eqnarray}
with $j$ and $\tablo{\nu}$ as specified in (\ref{j-cond}) and
(\ref{nuprime}).

\subsection{Norms for the fqH basis}

Of importance for later calculations are the norms of the
states $|\{m_j ; n_i\}\rangle^Q$ of the fqH basis. These norms
can be evaluated by using the fqH-CS correspondence, together with
the norms in the CS basis, as specified in (\ref{jnorm})
and (\ref{phinorm}). The general result is
\begin{equation}
N_{\{m_j;n_i\}}=  
\left| |\{ {m_j ; n_i}\rangle^Q \right|^2
=( \jip{\nu'}{p})^2\jip{\sigma'}{{{1\over p}}} \ ,
\end{equation}
where $\tablo{\nu}$ and $\tablo{\mu}$ are the tableaus corresponding
to $\tablo{n_i}$ and $\tablo{m_j}$, respectively, and
the tableau $\tablo{\sigma}$ is specified in (\ref{fqHtoCS}).
The norm can be
factorized into the norms of the $J$ and $\phi$ parts separately
times an extra factor associated with the added partition $N^M$ 
\begin{equation}
N_{\{m_j;n_i\}}=
 \jip{\nu'}{p}\jip{\mu'}{{1\over p}}\prod_{(i,j)\in M^N} 
 { (\mu_j-i+M)+p(\nu_i-j+N+1)\over (\mu_j-i+1+M)+p(\nu_i-j+N)} \ .
\end{equation}
The expressions (\ref{2partnorms}) are special cases of this general
formula.

In the case where $m_j, n_i \gg 1$
the expressions simplify and one finds the following factorized form
\begin{equation}
N_{\{m_j;n_i\}} \approx
\prod_{j=1}^M { {m_j}^{p-1}  \over \Gamma(p)} \,
\prod_{i=1}^N { {n_i}^{{1 \over p}-1}  \over \Gamma({1 \over p})} \ .
\label{normasymp}
\end{equation}

\section{Form factors}
\setcounter{equation}{0}
\setcounter{figure}{0}

\subsection{Vacuum form factors}

We start by considering the simplest non-vanishing form
factors of the basic electron operators $J(z)$, $J^\dagger(z)$
against the multi-particle states in the fqH basis
\bea
&&  \bra 0 | J_{{p \over 2}+m} | \{ n_p, \ldots , n_2 , n_1 \} \ket_N
    = [N_{ \{ n_p, \ldots , n_2 , n_1 \}}]^{1 \over 2}
      f_J(n_p,\ldots,n_1) \, \delta_{m,n_p+\ldots+n_1} \ ,
\nonu
&&   \bra 0 | J^\dagger_{+{p \over 2}+m} | \{ m_1 \} \ket_N
    = [N_{\{m_1\}}]^{1 \over 2} \, f_{J^\dagger}(m_1) \, \delta_{m,m_1} \  ,
\label{ffdef}
\eea
where the subscript $N$ indicates that the state has been
properly normalized.

One immediately finds
\be
f_{J^\dagger}(m_1) = 1 \ .
\ee

We briefly explain the exact evaluation of the form
factor $f_J(n_p,\ldots,n_1)$ as defined in (\ref{ffdef}).
Let us consider the special case $p=2$ first. In that
case the operator $J(z)$ has conformal dimension
$1$ and may be identified with one of the currents
of the affine Kac-Moody algebra $\widehat{su(2)}_1$
(see appendix A). By exploiting the OPE
\be
\phi(w_1) \phi(w_2) =
  (w_1-w_2)^{+\half} \left[ J^\dagger(w_2)
  + {\cal O}(w_1-w_2) \right]
\ee
one obtains
\be
J^\dagger(w_2) = \oint_{C_{w_2}} {dw_1 \over 2\pi i}
  (w_1-w_2)^{-{3 \over 2}} \phi(w_1) \phi(w_2) \ .
\ee
Using the expansion formula (\ref{phiexpansion}) we obtain
\be
J^\dagger(w_2) | \, 0 \, \rangle =
\sum_{n_2,n_1} P^{{1 \over 2}}_{ \{n_2,n_1\}}(w_2,w_2)
| \{n_2,n_1\}_\phi, q=2 \rangle
\ee
and it follows that
\be
 \bra 0 | J_{1+m} \, | \{n_2,n_1\} \ket_N
 =  [N_{\{n_2,n_1\}}]^{1 \over 2}  \,
    P^{1 \over 2}_{\{n_2,n_1\}}(1,1) \, \delta_{m,n_2+n_1} \ .
\ee
For general $p$ one obtains a similar result in terms of Jack
polynomials with label ${1 \over p}$. Using explicit expressions
\cite{St,LPS} for $P_{\{\ldots\}}^{1 \over p}(1,\ldots,1)$, we
obtain
\be
f_J(n_p,\ldots,n_1) =
    \prod_{i=0}^{p-1} {\Gamma({1 \over p}) \over
                       \Gamma(1-{i \over p})}
    \prod_{i<j} {\Gamma(n_j-n_i+ {j-i+1 \over p}) \over
                 \Gamma(n_j-n_i+ {j-i \over p})} \ .
\ee
In the large volume limit, where all $n_i\gg 1$, this
becomes
\be
f_J(n_p,\ldots,n_1) =
  {[\Gamma({1\over p})]^p \over
   \prod_{i=0}^{p-1} \Gamma(1-{i \over p})}
   \, \prod_{i<j} (n_j-n_i)^{{1 \over p}} \ .
\label{jastrow}
\ee

The form (\ref{jastrow}) of the form factor can be
viewed as a limit in (chiral) CFT of a result on correlation
functions for the `classical' model of quantum mechanics
with inverse square exchange. This result was conjectured
by Haldane \cite{Hal3} and later proven in \cite{Ha,LPS}.

An important insight is that there are no other
non-vanishing form factors of the state
$J^\dagger_{-{p \over 2}-m}\vac $ with the elements of
the fqH basis. In other words, the spectral weight of
this state is completely accounted for by
states having precisely the minimal number of $p$
quasi-holes. On the basis of this observation, it has
been proposed that the $T=0$ particle system underlying
the fqH basis be viewed as an `ideal gas of fractional
statistics particles'. In formula, this completeness is
expressed by the following identity for the $T=0$ 2-point
function of the edge electron operator $J_{-t}$
\bea
{1 \over m+1} \lefteqn{\bra 0 | J_{m+1} J^\dagger_{-m-1} | 0 \ket}
\nonu
&&
= {1 \over m+1} \sum_{n_2\geq n_1, n_2+n_1=m}
  \bra 0 | J_{m+1} | \{n_2,n_1\} \ket_N
  {}_N\bra \{ n_2, n_1 \} | J^\dagger_{-m-1} | 0 \ket
\nonu
&& = {1 \over m+1} \sum_{n_2\geq n_1, n_2+n_1=m}
     {(n_2-n_1) \over (n_2 n_1)^{1 \over 2}} 
  = \int_0^{1 \over 2} dx {(1-2x) \over [x(1-x)]^{\half} } = 1 \ ,
\eea
where $x={m_1 \over m}$ and we inserted asymptotic expressions
valid for $m,n_2,n_1\gg1$. This result is in agreement with a
direct computation using algebraic properties of the $J$ modes.

We remark here that, as we shall see in the next sections,
the structure of more general form factors, shows
`many body effects' and is not easily reconciled with a
notion of an ideal gas of fractional statistics particles.

\subsection{More general form factors}

We now consider more general form factors for the
edge electron `annihilation operator' $J^\dagger(z)$.
The simplest form factor with a 2-particle in-state is
\bea
\lefteqn{
  {}_N \bra \{m_1^\prime\} |
  J^\dagger_{{3p \over 2}+m} | \{ m_2,m_1 \} \ket_N = }
\nonu &&
    \left[ N_{\{m_2,m_1\}} \over N_{m_1^\prime} \right]^{1 \over 2}
    \, f_{J|JJ}(m_1^\prime,m_2,m_1) \, \delta_{m,m_2 + m_1 - m_1^\prime} \ .
\label{FFjjj}
\eea
Using the expansion formula (\ref{Jexpansion}) we derive the
following general result
\be
f_{J|JJ}(m_1^\prime,m_2,m_1) =
\sum_{s=0}^p C_s^{(p)} p^{(p)}_{m_1^\prime-m_1-s}
\ee
where
\be
P^p_{\{m_2,m_1\}}(z_1,z_2) = z_1^{m_2} z_2^{m_1}
 \sum_{l=0}^{m_2-m_1} p^{(p)}_l \left( {z_2 \over z_1}\right)^l \ .
\ee
Specializing to $p=1,2,3$, we have the following explicit results
\bea
\lefteqn{
p=1: \qquad f_{J|JJ}(m_1^\prime,m_2,m_1) =
\delta_{m_1^\prime =m_1} - \delta_{m_1^\prime =m_2+1}}
\nonu
\lefteqn{
p=2: \qquad f_{J|JJ}(m_1^\prime,m_2,m_1) = }
\nonu &&
\delta_{m_1^\prime =m_1} + \delta_{m_1^\prime =m_2+2} 
             - { 2 \over m_2 - m_1 +1}
             \Theta(m_1 < m_1^\prime < m_2+2) 
\nonu
\lefteqn{
p=3: \qquad f_{J|JJ}(m_1^\prime,m_2,m_1) = }
\nonu &&
\delta_{m_1^\prime =m_1} - \delta_{m_1^\prime =m_2+3}
               - { 6 (m_2+m_1-2m_1^\prime) \over
                   (m_2 - m_1 +1)(m_2 - m_1 +2)}
               \Theta( m_1 < m_1^\prime < m_2+3) \ .
\nonu
\label{FFjres}
\eea

We also consider the case where the in-state contains one $\phi$
and one $J$ quantum
\be
    {}_N \bra \{n_1^\prime\} |
    J^\dagger_{{p \over 2}+m} | \{ m_1; n_1 \} \ket_N =
    \left[ N_{\{m_1;n_1\}} \over N_{n_1^\prime} \right]^{1 \over 2}
    \, f_{\phi|J\phi}(n_1^\prime,m_1,n_1) \,
    \delta_{m,m_1+n_1 - n_1^\prime} \ .
\label{FFphijphi}
\ee
Using the expansion formula
\be
:J(z)\phi(w): | 0 \ket =
\sum_{m_1;n_1} r_{m_1,n_1}(z,w) | \{m_1;n_1\} \ket
\ee
with
\be
r_{m,n}(z,w) = z^{m+1}w^n - { m+1+ pn \over m+p(n+1)}z^mw^{n+1} \ ,
\ee
we derive
\be
f_{\phi|J\phi}(n_1^\prime,m_1,n_1) =
\left[ \delta_{n_1^\prime = n_1}
             - { p-1 \over m_1+p(n_1 +1)}
             \Theta(n_1^\prime > n_1) \right] \ .
\label{FFphires}
\ee

While the results for $p=1$ are a direct consequence of the
Wick theorem, the expressions for $p\neq 1$ show that
the `ideal gas interpretation' is no longer applicable for
general $p$: both form factors (\ref{FFjjj}) and (\ref{FFphijphi})
can be non-vanishing when the energy of the electron annihilation
operator $J^\dagger$ does not match the incoming electron energies
$m_2$ or $m_1$, and where the energy difference is transferred to
a second `spectator particle'.
Furthermore, $f_{J|JJ}$ and $f_{\phi|J\phi}$
are not the only non-vanishing
form factors of $J^\dagger$ with two incoming particles. For example,
there are non-vanishing overlaps between a state
created by applying  $J^\dagger$ on the a 2-electron state
and states containing more quasi-particles than just a single
electron. The additional quasi-particles can be visualized
as (neutral) density waves or excitons, which are composed
of a single electron and $p$ quasi-holes. In the next sub-section,
we explore the selection rules that determine in a more
general setting the possible out-states for which the
form factor of $J^\dagger$ with a given in-state is non-vanishing.

\subsection{A form factor selection rule}

In this section we put a bound on the possible out-state
that arise upon acting with an electron creation or annihilation
operator on a given in-state. We perform this analysis in
the CS basis, where the systematics of Jack polynomials come to
our help. Using the mapping of section 3.4, the results can be
translated to the fqH basis.

We focus on the form factors
\begin{equation}
  \langle \tablo{\mu}_J,{q}| J_{-m+q+p-{p \over 2}}
  |\tablo{\nu}_J,q+p\rangle \ ,
\label{formfac}
\end{equation}
with (unnormalized) Jack states as in (\ref{jackstate}).

The power of Jack Polynomial technology in
analyzing this form factor comes from the fact a product of
vertex-operators can be written as a sum over products of a
`coordinates' Jack polynomials $\Pjack{\lambda}{p}{\{z_i\}}$ and
`bosonic modes' Jack operators $\JEjack{\lambda}{{1\over p}}$. Both
the coordinate and the bosonic mode Jack polynomials can be
manipulated using the results for Jack polynomials in mathematical
literature \cite{St,McD}. In Appendix B, we used this to rearrange
the part of the vertex operators that survives after applying them
to the vacuum.  Here we apply the same method which
enables us to analyze the action of a single mode operator
$J_{-m+q+p-{p \over 2}}$ on a state created by a Jack operator.
We note that for $m\geq 0$ this mode operator creates an
additional edge electron.

A product of $N+1$ edge electron operators acting on a
vacuum $|\tilde{q}\rangle$ can be expanded in Jack polynomials and 
Jack operators,
\begin{eqnarray}
  \lefteqn{J(w)\prod_{i=1}^N J(z_i) |{\tilde{q}}\rangle=}\nonu
  && w^{-\tilde{q}}\prod_{i=1}^N(w-z_i)^p
  \prod_{i<j}(z_i-z_j)^p\prod_{i=1}^N z_i^{-\tilde{q}}
  \nonu
&&\times \left(\sum_{\{\lambda\}}(-1)^{|\lambda|}
    \Pjack{\lambda}{p}{w}\JEjack{\lambda'}{{1\over p}}\right)
  \left(\sum_{\{\rho\}}
    (-1)^{|\nu|}\Pjack{\rho}{p}{\{z_i\}}\JEjack{ \rho' }{{1\over p}}\right)
  |{\tilde{q}}-pN-p \rangle.\nonumber\\
  \label{eq:JackExp}
\end{eqnarray}
The state $|\tablo{\nu}_J, {\tilde{q}}-Np \rangle$ with $N=l(\tablo{\nu})$
can be extracted from $\prod_{i=1}^N J(z_i)
|{\tilde{q}}\rangle$ by applying the operator 
\begin{eqnarray}
  \lefteqn{\Ojack{\nu}{\{z_i\}}{\tilde{q}} =} \nonu
  &&(\pip{\nu}{p}{l(\tablo{\nu})})^{-1} 
  \left( \prod_{i=1}^{l(\tablo{\nu})}\oint{dz_i\over  2\pi i}
  {1\over  z_i^{-\tilde{q}+1}}\right)
  {\Pjack{\nu}{p}{\{ z_i^{-1}\}}}
\Delta^p\left(\left\{{ z_i^{-1}}\right\}\right) \ , 
\nonu
\end{eqnarray}
where we made use of the inner product (\ref{eq:altip}) on coordinate
dependent Jack polynomials. The norm  $\pip{\nu}{p}{l(\tablo{n}}$  of 
the Jack polynomials will drop out of the final result. We can write
\begin{eqnarray}
  \lefteqn{J_{-m+q+p-{p \over 2}}|\tablo{\nu}_J,\tilde{q}-pl(\tablo{\nu})
\rangle =}\nonu
&& \Mmode{m-q-p}{w}
  J(w) \Ojack{\nu}{\{z_i\}}{\tilde{q}}
  \prod_{i=1}^{l(\tablo{\nu})} J(z_i) |{\tilde{q}} \rangle
\end{eqnarray}
where
\begin{equation}
\Mmode{m}{w}= \oint{dw\over  2\pi i} {1\over  w^{m+1}} \ .
\end{equation}

{}From the definition it is clear that $\Ojack{\nu}{ \{z_i\} }
{\tilde{q}}$ commutes with $J(w)$, which allows us to
interchange the order in the expression above and to use the
expansion (\ref{eq:JackExp}) of $J(w)\prod_{i=1}^N
J(z_i) |{\tilde{q}}\rangle$ in terms of Jack polynomials.
By taking the inner product of the resulting expression with 
$\langle\tablo{\mu}_J, q |$, where $q=\tilde{q}-(N+1)p$, 
we obtain an expression for the form factor (\ref{formfac}).

To decide for which choices of $\tablo{\mu}$, $\tablo{\nu}$
and $m$ the form factor can be non-vanishing we proceed as
follows. We first rewrite the product
\begin{equation}
  \prod_{i=1}^N(w-z_i)^p =
  \sum_{\{n_i\}}C_{\{n_i\}}^{(p)} w^{p N}\prod_{i=1}^N \left({z_i\over
w}\right)^{n_i}
\end{equation}
where $n_i=0,1,\dots ,p$. We insert this into the expansion given in
(\ref{eq:JackExp}), and we use that $\Pjack{\lambda}{p}{w}$ is zero
when $\tablo{\lambda}$ is not of the form
$\tablo{\lambda}=(\lambda_1,0,0,...)$. The contour integration
contained in $\Mmode{m-p-q}{w}$ selects the following value for
$\lambda_1$
\begin{equation}
\lambda_1 =|n|+m \ .
\end{equation}
Varying $|n|=\sum_i n_i$ over all allowed values $|n|=0,\dots, pN$, we
find that $\lambda_1$ has to satisfy the inequalities
\begin{equation}
  m \leq \lambda_1 \leq m+pN \ .
\label{lambdaineq}
\end{equation}

We write 
$\langle \tablo{\mu}_J| \tablo{\lambda}_J\tablo{\rho}_J\rangle$ 
for the inner product
$\langle \tablo{\mu}_J,q| \JEjack{\lambda'}{{1\over p}}
\JEjack{\rho'}{{1\over p}}|q\rangle$. 
Using a result by Stanley (Proposition 5.3 from \cite{St}) we learn 
that this inner product is non-zero if and only if $\rho \subseteq \mu$ 
and $\mu/\rho$ is a horizontal $\lambda_1$-strip.
The relation $\rho \subseteq \mu$ indicates that for all $i$ we have
$\rho_i\leq \mu_i$. The skew tableau $\mu/\rho$ is the tableau
containing all boxes which are in the tableau  $\mu$ but not in the
tableau $\rho$. If every column of the skew tableau contains at most one
box it is called a horizontal strip and if furthermore  the total
number of boxes in it is $\lambda_1$ it is called a horizontal 
$\lambda_1$-strip.

A particular consequence is that $\lambda_1$ satisfies 
\be
  0\leq\lambda_1\leq\mu_1 \ .
\ee 
Combining this inequality with (\ref{lambdaineq}), we conclude that
\begin{equation}
\max(0,  m)  \leq \lambda_1 \leq \mu_1 \ ,
\end{equation}
which implies that 
\begin{equation}
\mu_1< m  \quad \Rightarrow \quad
J^\dagger_{m-q-p+{p\over2}}|\{\mu\}_J,q\rangle =0 \ ,
\end{equation}
in agreement with explicit step-functions in the
form factors (\ref{FFjres}) and (\ref{FFphires}).

The form factor is now written as
\begin{eqnarray}
\lefteqn{ {} \langle q , \tablo{\mu}_J
  |J_{-m+q+p-{p \over 2}}|\tablo{\nu}_J,{q}+p\rangle =}\nonu
  && (\pip{\nu}{p}{l(\tablo{\nu})})^{-1}
  \sum_{\{n\},\{\lambda_1\},\{\rho\}}
  \delta_{|n|+m,\lambda_1} \,
  C_{\{n\}}^{(p)} \, 
  \langle\tablo{\mu}_J|\tablo{\lambda_1}_J\tablo{\rho}_J\rangle
  \nonumber\\ 
  &&\times \left( \prod_{i=1}^N \Mmode{0}{z_i} \right)m_{\{n\}}(\{z_i\})
  \Delta^p\left(\left\{{ z_i^{-1}}\right\}\right) \Delta^p(\{z_i \})
   \Pjack{\rho}{p}{\{z_i\}}
  \Pjack{\nu}{p}{\left\{{ z_i^{-1}}\right\}} ,
\nonumber\\
\label{eq:theorem}
\end{eqnarray}
where the summations extend over $i,j=1,\dots, N$ and
$\{n\}=(n_1,\dots,n_N)$ with $n_i\in{0,1,\dots,p}$. 

\setlength{\unitlength}{1 em}
\begin{figure}[t]
  \begin{picture}(15,17)(-10,0)
    \put(1,15){\framebox(1,1){}}
    \put(5,15){\framebox(1,1){}}
    \multiput(9,15)(1,0){9}{\framebox(1,1){}}
    \put(1,14){\framebox(1,1){}}
    \put(3,14.5){\makebox(1,1){\dots}}
    \put(5,14){\framebox(1,1){}}
    \multiput(5,12.5)(0,-0.5){3}{\makebox(1,1){$\cdot$}}
    \put(7,14.5){\makebox(1,1){$\dots$}}
    \multiput(9,14)(1,0){7}{\framebox(1,1){}}
    \put(1,10){\framebox(1,1){}}
    \put(5,10){\framebox(1,1){}}
    \put(7,10){\makebox(1,1){$\dots$}}
    \multiput(9,10)(1,0){2}{\framebox(1,1){}}
     \multiput(1,8.5)(0,-0.5){3}{\makebox(1,1){$\cdot$}}
    \multiput(5,8.5)(0,-0.5){3}{\makebox(1,1){$\cdot$}}
   \multiput(1,6)(0,-1){5}{\framebox(1,1){}}
     \put(3,10){\makebox(1,1){$\dots$}}
    \multiput(5,6)(0,-1){2}{\framebox(1,1){}}
    \put(1,15){\makebox(1,1){1}}
    \put(5,15){\makebox(1,1){$p$}}
    \put(1,14){\makebox(1,1){2}}
    \multiput(1,12.5)(0,-0.5){3}{\makebox(1,1){$\cdot$}}
    \put(1,10){\makebox(1,1){$l$}}
  \end{picture}
  \caption{$l$ arms and $p$ legs.}
  \label{fig:armslegs}
\end{figure}

The last part of this expression is the inner product on 
products of Jack polynomials with a finite number of arguments 
$\{z_i\}$. In Appendix C we discuss restrictions on the tableaus
$\{n\}$, $\tablo{\nu}$ and $\tablo{\rho}$ that follow from
imposing that this final inner product be non-zero. 
Combining all ingredients, one arrives at the following 
\begin{quote}
\noindent {\bf Form factor selection rule}

\noindent
{\it The form factor 
$$
  \langle \tablo{\mu}_J,q| J_{-m+q+p-{p \over 2}}|
  \tablo{\nu}_J,q+p\rangle
$$               
can only be non-zero if $\tablo{\mu}$, $\tablo{\nu}$, $m$            
satisfy the following conditions
\begin{itemize}
\item[a.]  $|\nu|+m =|\mu|$
\item[b.]
  \begin{itemize}
  \item[1.]  $\nu_j \geq \mu_{j+1}$ for all $j$
  \item[2.]  $\nu_i \leq p$ for $i>l(\tablo{\mu})$
  \end{itemize}
\item[c.]  $m+\sum_{i \geq l(\tablo{\mu})}\nu_i\leq \mu_1$
\item[d.]  $\sum_{i=1}^{j}\nu_i  \leq \sum_{i=1}^{j}(\mu_{i}+p )$.
\end{itemize}
These conditions imply that the tableau $\tablo{\nu}$ should have at
most $p$ legs and $l(\tablo{\mu})$ arms, see fig.~\ref{fig:armslegs}.}
\end{quote}
We refer to Appendix C for a complete proof of this result.

We remark that the above selection rule can be viewed as a 
generalization of a selection rule that was used by Lesage, Pasquier 
and Serban \cite{LPS} for the evaluation of the zero-temperature 
density-density correlation function in the Calogero-Sutherland model.
These authors found that the (neutral) density operator $\rho$ when 
acting on the vacuum creates an `exiton' with $p$ quasi-holes and
a single electron, corresponding to a Young tableau with $p$ legs
and a single arm. In the processes described by the form factor 
discussed in this section a similar structure is found. Starting from 
a multi-$J$ state described by a tableau $\tablo{\mu}$, the operator
labeled by $m$ annihilates one of the $J$-quanta. If there
is a mismatch between the modes of the operator and of the quantum 
that is annihilated, the remaining momentum is carried away by a 
density fluctuation, which roughly speaking corresponds to one
extra arm and the $p$ legs that can be present in the tableau 
$\tablo{\nu}$. If one starts from a state which, in the fqH basis,
has a number of $J$-quanta and $N$ quasi-holes, with maximal mode 
$n_N$, one finds that upon annihilating a $J$ mode up to $n_N+1$ 
exitons can be created.

\subsection{Relation with $S$-matrix approach}

In this section, we consider the structure of the
quasi-particle form factors from the point of view of an
associated $S$-matrix structure.

Via the TBA procedure, the distribution functions
for fractional exclusion statistics
are linked to an $S$-matrix
with the following dependence on particle rapidities
$\theta=\theta_2-\theta_1$
\be
S_{ab}(\theta) = \exp[2\pi i (\delta_{ab}-G_{ab})
\Theta(\theta)] \ .
\ee

Although the quasi-particle states that we have
considered are part of the discrete spectrum of
a finite size system, it is natural to identify
the quasi-particle states with a set of asymptotic
particle states in a scattering theory with
2-body $S$-matrix of this type, with diagonal
statistics matrix $G_{11}=p$, $G_{22}={1 \over p}$.

Via the well-known form factor axioms, this identification
leads to specific properties of the form factors. In particular,
we expect factors
\be
(\eps_i - \eps_j)^p \ ,
\qquad
(\tilde{\eps}_i - \tilde{\eps}_j)^{1 \over p} \ ,
\ee
in form factors with particles $J(\eps_i)$ and
$\phi(\tilde{\eps}_j)$ in the in-state, and
annihilation poles between particles in the
in- and out-states. 

The explicit result
(\ref{jastrow}) for the vacuum form factor 
$f_J$ has the expected zero's 
$(n_i-n_j)^{1 \over p}$. For the more general form 
factors discussed in section 4.2 the structure is less 
clear. We observe however that, upon heuristically 
replacing
\be
\delta_{m_2,m_1} \rightarrow {1 \over (\eps_2-\eps_1)} \ ,
\qquad
\Theta(m_2 - m_1) \rightarrow \log(\eps_2-\eps_1)
\ee
we have (for $p=1,2,3$)
\be
[ \del_{\eps^\prime_1} ]^{p-1}
\left[ f_{J|JJ}(\eps^\prime_1, \eps_2,\eps_1) \right]
\rightarrow
 { (\eps_2 - \eps_1)^p \over
   (\eps^\prime_1-\eps_2)^p (\eps^\prime_1-\eps_2)^p } \ .
\ee
It will be most interesting to investigate whether
the asymptotic limit of the form factors considered
and computed in this paper can be obtained by means
of an axiomatic approach.

\section{Form factor expansion at finite temperature}
\setcounter{equation}{0}
\setcounter{figure}{0}

\subsection{General remarks}

In a system of non-interacting electrons, transport
properties such as $I$-$V$ and noise characteristics
are obtained by computing the relevant amplitudes
for transmission and reflection of single particles, and
then performing a statistical average using a one-particle
Fermi-Dirac distribution function. An important goal,
that we had in mind when setting up the quasi-particle
formulation of fqH edges, is to arrive at a similar
description of transport processes in these interacting,
highly non-Fermi liquid, systems.

As a first attempt in this direction, one may try to
simply replace free electron amplitudes by corresponding
amplitudes for fqH quasi-particles, and simultaneously
replace the Fermi-Dirac distribution by an appropriate
distribution function for fractional statistics. While,
as we shall argue, this idea is essentially correct,
we stress that a correct implementation is subtle and
involves the important concept of a so-called form factor
expansion.

In this section, we shall focus on the following finite
temperature Green's functions in the CFT for the
$\nu={1 \over p}$ fqH edge
\be
h(\epsilon)=
\langle \psi^\dagger_{\nu={1 \over p}}(\eps)
        \psi_{\nu={1 \over p}}(\eps) \rangle_T \ ,
\qquad
H(\epsilon)=
\langle \psi_{\nu={1 \over p}}(\eps)
        \psi^\dagger_{\nu={1 \over p}}(\eps) \rangle_T \ ,
\label{green1}
\ee
where the operators $\psi^\dagger_{\nu={1 \over p}}(\eps)$
and $\psi_{\nu={1 \over p}}(\eps)$ are the continuum limits
of the edge electron operators $J_s$ and $J_s^\dagger$
considered in this paper.

In the next subsection, we recall how this Green's
function is used for the computation of the $I$-$V$
characteristics for the tunneling of electrons
into a fqH edge. After that, we give the general form
of the form factor expansions for finite temperature
correlation functions. We shall then zoom in on the
case $m=2$, and explain how the finite-$T$ Green's function
$h(\epsilon)$ can be approximated in a form factor
expansion.

\subsection{Kinetic equation for electron tunneling}

As explained in \cite{We2,vES}, the Green's functions
(\ref{green1}) can be
used to computed the finite temperature $I$-$V$ characteristics
for the tunneling of electrons from a Fermi Liquid (FL) reservoir
into a $\nu={1 \over p}$ fqH edge. Starting from the tunneling
hamiltonian
\be
H_{int} \propto t \, \int d\eps \,
\left[ \psi_{{\rm FL}}^{\dagger}(\eps)
       \psi_{\nu={1\over 3}}(\eps)
       + {\rm h.c.} \right] \ ,
\ee
one can show that, in lowest order perturbation theory, the
current-voltage characteristics are given by
\be
I(V,T) \propto e \, t^2 \int_{-\infty}^\infty d\epsilon
\left[ f(\epsilon-eV)H(\epsilon)-F(\epsilon-eV)h(\epsilon) \right] \ ,
\label{current}
\ee
where $f(\eps)$ and $F(\eps)$ are the standard Fermi-Dirac 
distributions for particles and holes.
Using the conformal mapping from a plane to a cylinder, or
employing an imaginary time approach, one finds the following
exact expression for the case $\nu={1 \over 3}$
\be
H(\epsilon)={ \epsilon^2+{\pi^2\over \beta^2} \over
                       e^{-\beta \epsilon }+1 }
\ , \quad
h(\epsilon)={ \epsilon^2+{\pi^2\over \beta^2} \over
                      1+e^{\beta \epsilon} } \  .
\label{green2}
\ee
They lead to $I$-$V$ characteristics
\be
I(V,T)\propto e \, t^2 \, \beta^{-3} \left({\beta eV\over 2\pi}+
\left({\beta eV\over 2\pi}\right)^3\right) \ ,
\label{ivchar}
\ee
in agreement with the result obtained in different approaches
\cite{KF,dCF}. The $I$-$V$ characteristics (\ref{ivchar})
show cross-over from a linear (thermal) regime into a power-law
behavior at high voltages and thus presents a clear fingerprint
of the Luttinger liquid features of the fqH edge. The experimental
results of \cite{CPW} are in agreement with these predictions.
(See \cite{dCF} for a further theoretical analysis of the data.)

\subsection{Form factor expansion}

As a proto-type study for a form factor expansion
based on CFT quasi-particles, we now analyze the
Green's function $h(\eps)$, for $p=2$ in that spirit.
Obviously, an exact result is easily obtained
\be
h(\epsilon) = {\epsilon \over e^{\beta \epsilon} -1} \ .
\ee
The Bose-Einstein denominator in this expression has
its origin in the fact that the operators $J$, $J^\dagger$
satisfy bosonic commutation relations. In
the spirit of the quasi-particle formulation of this
paper, we wish to treat the $J$, $J^\dagger$-quanta as
quasi-particles with exclusion statistics $g=2$, and see if
we can recover the Green's function $h(\epsilon)$ in
such an approach.

The Green's function $h(\epsilon)$ can be viewed as a
one-point function for the operator
$N_{\psi}(\epsilon) =
\psi^\dagger_{\nu={1 \over p}} \psi_{\nu={1 \over p}}(\eps)$.
In the formulation on the finite system of size $L$,
this operator is represented as
$N_J(m) = a J_{-m} J^\dagger_m$, with
$\epsilon = a m$, with $a ={2 \pi \over L \rho_0}$
the energy level spacing in the finite size system.
This one-point function is formally expressed as
\be
 { \sum_{\Psi \in {\cal H} }
   \bra \Psi  | N_J(m) | \Psi \ket
 \over
   \sum_{ \Psi \in {\cal H} }
   \bra \Psi | \Psi \ket} \ .
\ee

The sum runs over a basis the full Hilbert space of the edge
CFT, and we can opt for the fqH quasi-particle basis
discussed in this paper. The idea is now that the matrix
elements $\bra \Psi | N_J(m) | \Psi \ket$ are dominated by 
processes where only a few of
the quasi-particles that are present in a concrete basis
state $| \{ m_i;n_j \} \ket$ participate (we restrict our 
attention to states in the $Q=0$ sector of the fqH basis).

For the case at hand, the lowest contributions comes from
1-particle states $| \{m_1\} \ket$, for which one computes
the form factor
\bea
\lefteqn{ D^{(1,0)}(m;m_1) =
  {}_N \bra \{m_1 \} | J_{-1-m}
   J^\dagger_{+1+m} | \{m_1\} \ket_N = }
\nonu &&
(m+1)\delta_{m,m_1} + 2 \left( 1 - {m+1 \over m_1+1} \right)
\Theta(m < m_1) \ .
\label{D10}
\eea
The expected presence of an edge electron of energy
$m_1$ is given by the distribution function
$\bar{n}_2(\epsilon_1=a m_1)$. This leads to the following
contribution to the Green's function
\be
h^{(1,0)}(\eps)
= a \sum_{m_1}
   D^{(1,0)}(m,m_1)
   \bar{n}_2(a m_1) \ .
\label{h10}
\ee
If we now consider the form factor of $N_J(m)$ against
a two-electron state, we find (see next subsection) that
it is not simply the sum of two 1-particle contributions.
The left-over part is what we call the irreducible
2-electron form factor
\bea
\lefteqn{ D^{(2,0)}(m;m_1,m_2) =
  {}_N \bra \{ m_1,m_2 \} | J_{-3-m}
     J^\dagger_{+3+m} | \{ m_2,m_1\}\ket_N }
\nonu &&
  - {}_N \bra \{m_1\} | J_{-3-m}
   J^\dagger_{+3+m} | \{m_1\}\ket_N
  - {}_N\bra \{m_2\} | J_{-3-m}
   J^\dagger_{+3+m} |\{ m_2\}\ket_N \ .
\eea
It leads to an additional contribution $h^{(2,0)}(m)$
to the Green's function
\be
h^{(2,0)}(\eps) 
 = a \sum_{m_1,m_2}
   D^{(2,0)}(m;m_1,m_2)
   \bar{n}_{2}(a m_1) \bar{n}_2(a m_2) \ .
\label{h20}
\ee
Similarly, we define
\bea
\lefteqn{D^{(1,1)}(m;m_1,n_1)=}
\nonu &&
{}_N\bra \{n_1,m_1\}|J_{-1-m}J^\dagger_{+1+m}|\{ m_1,n_1\}\ket_N
- {}_N\bra \{m_1\}|J_{-1-m}J^\dagger_{+1+m}|\{ m_1\}\ket_N
\nonu
\eea
and
\be
h^{(1,1)}(\eps)
 = a \, \sum_{m_1,n_1}
   D^{(1,1)}(m;m_1,n_1)
   \bar{n}_2(a m_1) \bar{n}_{1 \over 2}(a n_1) \ .
\label{h11}
\ee
Continuing in this manner, we build up the following expansion
\bea
&&
h = \sum_{M,N} h^{(M,N)}(\eps) \ ,
\nonu
&&
h^{(M,N)}(\eps) = a \, \sum_{\{m_i;n_j\}}
  D^{(M,N)}(m; \{m_i;n_j\})
  \prod_i \bar{n}_2(a m_i)
  \prod_j \bar{n}_{1 \over 2}(a n_j) \ .
\nonu
\label{FFexp}
\eea

We remark that an expansion of precisely this type
has been proposed by LeClair and Mussardo \cite{LM},
see also \cite{Sa}. This work was
done in the context of integrable qft's, that are fully
characterized by a factorized $S$-matrix. In such
a context, the irreducible form factors are
constrained by the form factor axioms, and the distribution
functions have their origin in a TBA procedure. Although
clearly in the same spirit, the analysis that we present
here is very different at the technical level. We obtain
the relevant form factor by explicit
computation in a theory that is regularized by
the finite size of the fqH edge, and we have identified
the relevant distribution functions by analyzing
the state counting of the (discrete) spectrum of the
finite-size system. We thus do not rely on an
underlying (massless) $S$-matrix point of view.
Despite these differences, it seems clear that the
two approaches are closely related: in subsection
4.5 we briefly indicated that our form factor have symmetry
properties that are expected on the basis of
a `purely statistical $S$-matrix'. We leave this
interesting issue for further study.

\subsection{Irreducible form factors}

To evaluate explicitly the leading terms in the
form factor expansion (\ref{FFexp}) for $h(\epsilon)$,
we need to evaluate the relevant irreducible
form factors. While it is clear that these
form factors have very special mathematical properties, we
here compute them by a simple brute force
computation, relying on the explicit form of the
two-particle states (\ref{jjstate}) and (\ref{jphistate}),
and on the algebraic properties of the operators
$J$, $J^\dagger$ and $\phi$ (see appendix A).

\subsubsection{Two electrons}

For the irreducible two electron form factor we find
\bea
\lefteqn{ D^{(2,0)}(m;m_2,m_1) = }
\nonu
&&
\delta_{m-m_2} \, {-2(m_2+3) \over m_2-m_1+3}
+
\delta_{m-m_1+2} \, {-2(m_1+1) \over m_2-m_1+1}
\nonu
&&      
+ {4\over (m_2-m_1+3)}{1\over (m_2-m_1+1)}
    {1\over (m_1+1)(m_2+3)}
\nonu &&
\quad \times   [\Theta(m<m_1-2) \, P(m; m_1,m_2)
\nonu &&
\qquad +\Theta(m<m_2<m+m_1) \, Q(m; m_1,m_2)
\nonu &&
\qquad +\Theta(m<m_2) \, R(m; m_1,m_2) ] \ ,
\eea
with
\bea
\lefteqn{ P(m; m_1,m_2) =}
\nonu &&
(m_2-m_1+3)(m_1-m-2)(2m_1-m_2-3)
\nonu &&
+(m_1-m-2)(m_1-m-3)(-3m_2 + {5 \over 3}m_1+{1 \over 3}m-{26 \over 3})
\nonu &&
+ (m+3) [-2(m_2-m_1+3)(2m_1-m-1)
\nonu &&       
 -2 m_1(m_1+1)+(m+3)(m_2+m_1-m+1)]
\nonu
\lefteqn{ Q(m; m_1,m_2) =}
\nonu &&
(m_1-m_2+m+1) [ (m_2-m_1+3)^2 +2(m_2-m_1+3)(m_1-m_2+m)
\nonu &&
+{2 \over 3}(m_1-m_2+m)(m_1-m_2+m-1)]
\nonu
\lefteqn{ R(m; m_1,m_2) =}
\nonu &&
(m_2-m)(m_1+1)(m_2-m_1+3) + {1 \over 3} m_1(m_1+1)(m_1+3m_2-3m+2) \ .
\nonu
\eea
The polynomials $P$, $Q$ and $R$ enjoy special properties,
which include
\be
(P+Q+R)(m; m_1,m_2) = -{1 \over 3}(m_1-m_2-1)(m_1-m_2-2)(m_1-m_2-3) \ .
\ee

\subsubsection{One electron and one quasi-hole}

The irreducible form factor with one electron and one hole
is found to be
\bea
\lefteqn{D^{(1,1)}(m;m_1,n_1)=}
\nonu &&
\delta_{m_1,m} \, {m_1+1 \over m_1+2n_1+1}
\nonu &&
+ \Theta(m<m_1)
  {1\over C_{n_1}^{(-{1\over 2})}(m_1+2n_1+2)(m_1+2n_1+1)(m_1+1)}
\nonu &&
  \quad \times \left[ C_{n_1-m_1+m}^{(-{1\over2})} \, S(m;m_1,n_1)
                + C_{n_1}^{(-{1\over2})} \, T(m;m_1,n_1) \right] \ ,
\eea
with
\bea
\lefteqn{ S(m; m_1,n_1) =}
\nonu &&
  (m_1+2n_1+1)^2 +
  (m+n_1-m_1) ({8\over 3}-4(m_1+2n_1+2)) +{4\over 3}(m+n_1-m_1)^2
\nonu
\lefteqn{ T(m; m_1,n_1) =}
\nonu &&
2(m_1-m)((m_1+2n_1+1)^2-1)
+ 2 (2n_1+1) (m_1-m-1)
\nonu &&
+ 2 ({2\over3}n_1+1) (2n_1+1) \ .
\eea

\subsection{Evaluating the series}

With the information collected in the previous subsections, we can
evaluate the 1-particle and 2-particle contributions
$h^{(1,0)}$, $h^{(2,0)}$ and $h^{(1,1)}$
to the Green's function $h(\epsilon)$.

The expressions (\ref{h10}), (\ref{h20}) and (\ref{h11}) for
$h^{(2,0)}$ and $h^{(1,1)}$
are discrete sums, which we wish to study in the limit
$a \to 0$. In this limit, one may view the
expressions as Riemann sums and evaluate them using continuous 
integrals; however, one needs to be careful because the integrands 
as they
stand are have singularities, and the sums are not term-by-term
convergent. One may check however that by carefully redistributing
some of the terms, one obtains convergent sums that can be
approximated by the corresponding continuous integrals.
Proceeding in this manner, and using a numerical integrator,
we obtained the results plotted in figure 5.1.

We observe that the form factor series converge in the following 
sense: while the 1-particle terms agree with the exact result for
${\eps}$ greater than about $3k_BT$, the approximation
including 2-particle terms reaches the exact curve at
${\eps}$ around $2k_BT$. For energies $\eps \ll k_BT$, the thermal 
factors do not efficiently suppress many particle contributions, 
and the convergence of the form factor expansion is expected to be 
slow.

We remark that the asymptotic behavior for $\eps\gg k_BT$
of the 2-particle terms is
\be
h^{(2,0)}(\eps) \sim c_2 e^{-\beta \eps} \ ,
\qquad
h^{(1,1)}(\eps) \sim c_{1 \over 2} e^{-\beta \eps}
\ee
with
\be
c_2 = -2 \int_0^\infty d \eps_1 \bar{n}_2(\eps_1) \ ,
\qquad
c_{1 \over 2}
= \int_0^\infty d \tilde{\eps}_1 \bar{n}_{1 \over 2}(\tilde{\eps}_1) \ .
\ee
Remarkably, the duality relation (\ref{dual}) leads to the relation
\be
c_2 = - c_{1 \over 2}
\ee
meaning that the Boltzmann tails of the 2-particle
terms precisely cancel. This `conspiracy' was needed
as, numerically, it is seen that the deviation between
the exact curve $h(\eps)$ and the 1-particle term
$h^{(1,0)}(\eps)$ is far smaller than the individual
Boltzmann tails of $h^{(2,0)}$ and $h^{(1,1)}$.

\section{Conclusions}
\setcounter{equation}{0}
\setcounter{figure}{0}

Summarizing the results collected in this paper, we
have made some first steps on the way to realizing
a computational scheme where the $T$-dependence
of physical observables in a fqH system (charge transport
properties in particular) is computed with direct reference to
fractional statistics of the fundamental quasi-particles.
We expect that on the basis of the formalism
presented here, meaningful claims about the observability
of the fractional statistics of CFT edge quasi-particles
can be formulated. We leave this most interesting
aspect for further study.

We remark the the continuum (CFT) limit of the CS model
provides an ideal testing ground for form factor expansions
for finite temperature correlation functions, such as discussed 
in section 5.3 and in the literature \cite{LM,Sa}. This is
because on the one hand the theory is explicitly regularized
by the finite extent of the spatial direction and, on the other,
the finite temperature Green's functions are known from
standard CFT considerations. 

We thank A.W.W.~Ludwig for many insightful comments and collaboration
in the early stages of this project, and F.H.L.~Essler and
F.A.~Smirnov for discussions. This research has benefited from 
the NATO Collaborative Research Grant SA.5-2-05(CRG.951303) and from 
support from the foundation FOM of the Netherlands. We thank
the Erwin Schr\"odinger Institute (Vienna) and the Centre
de Recherches Math\'ematiques (Montreal) for hospitality 
during the course of this work.

\appendix

\section{Algebraic properties of $\nu=\half$ edge operators}
\setcounter{equation}{0}
\setcounter{figure}{0}

The charged edge operators $J=J^-$, $J^\dagger=J^+$ in the
edge theory at $\nu=\half$ are part of a $SU(2)_1$ affine 
symmetry algebra. Together with the charge density 
$Q=i \sqrt{p}\del \varphi$ they satisfy
the commutation relations
\bea
&& [ J_{m_2}^+,J_{m_1}^- ] = m_1 \delta_{m_2+m_1} + Q_{m_2+m_1}
\nonu
&& [ Q_{m_2},J_{m_1}^\pm ] = \pm 2 J^\pm_{m_2+m_1} \ , \qquad
   [ Q_{m_2},Q_{m_1} ] = 2m_1 \delta_{m_2+m_1} \ .
\label{Jalg}
\eea
The fractionally charged edge quasi-particles $\phi^\pm$ transform 
in the spin-$\half$ representation of the $SU(2)$ symmetry
\be
[ J_m^\pm,\phi_s^\pm ] = 0 \ ,
\quad
[ J_m^\pm,\phi_s^\mp ] = \pm \phi_{m+s}^\pm \ ,
\quad
[ Q_m,\phi_s^\pm ] = \pm \phi_{m+s}^\pm \ .
\ee
Among themselves, the modes of $\phi^\pm$ satisfy
so-called generalized commutation relations, which 
have been studied in the context of the spinon formulation
of the $SU(2)_1$ CFT \cite{BPS,BLS}. 

\section{Jack polynomials and Jack operators}
\setcounter{equation}{0}
\setcounter{figure}{0}

In this appendix we briefly introduce the Jack polynomials
that are used in sections 3 and 4 of this paper. We essentially
follow the conventions of Iso~\cite{Is}, but we introduce 
different notations for the coordinate dependent Jack polynomials 
$\Pjack{\mu}{\beta}{\{z_i\}}$ and the
bosonic mode Jack operators $\Jjack{\mu}{\beta}{\left\{{a_{-n}\over 
\sqrt{\beta} }\right\}}$ . 
 
We start by specifying an inner product on the ring of symmetric
polynomials,
\begin{equation}
  \langle p_\tablo{\lambda}| p_\tablo{\mu} \rangle_\beta =
  \delta_{\tablo{\lambda},\tablo{\mu}}\beta^{-l(\tablo{\lambda})}z_\lambda
  \label{eq:psumip}
\end{equation}
where $p_\tablo{\lambda}(z_i)=  \prod_{j=1}^{l(\tablo{\lambda})}
  p_{\lambda_{j}}(\{x_i\})
  \;\;{\rm with}\;\;p_{\lambda_j}(\{x_i\})=\sum_i x_i^{\lambda_j}$ 
is the power sum set by a
Young tableau $\tablo{\lambda}=(\lambda_1,\lambda_2,\ldots, \lambda_l)$, 
$z_\tablo{\lambda}$ is $\prod_{i\geq1} i^{l_i}l_i!$ with $l_j$
the number of entries in $\tablo{\lambda}$ which satisfy $\lambda_i=j$
and $\beta$ a rational number. 

The coordinate Jack polynomials $\Pjack{\lambda}{\beta}{\{z_i\}}$ are
symmetric functions in the coordinates ${\{z_i\}}$ labeled by
a Young tableau $\{\lambda\}$ and a rational number $\beta$.
They are defined by the following properties
\\[2mm]
\noindent {orthogonality:
\samepage
\begin{center}
 $\langle\Pjack{\lambda}{\beta}{\{z_i\}} |\Pjack{\nu}{\beta}{\{z_i\}}
\rangle_\beta =\delta_{\tablo{\lambda},\tablo{\nu}} \jip{\nu}{\beta}
$ \\
\end{center}}
\noindent {triangularity:
\samepage
\begin{center}
  $\Pjack{\lambda}{\beta}{\{z_i\}}=\sum_{\tablo{\mu} }
v_{{\lambda}, {\mu}}(\beta) m_\tablo{\mu}$ where
$v_{{\lambda}, {\mu}}(\beta)=0$ unless $\tablo{\mu}\leq\tablo{\lambda}$ \\
\end{center}}
\noindent {normalization:
\samepage
\begin{center}
the coefficient $v_{{\lambda},{\lambda}}=1$ .
\end{center}}
\noindent
In this definition, $m_\tablo{\lambda}(\{z_i\})$ are the monomial
symmetric functions $\sum_{\sigma} \prod_i z_i^{\lambda_{\sigma(i)}}$
where $\sum_{\sigma}$ denotes the sum over all permutations of the
indices $i$. The partial ordering $\leq$ on partitions is the so-called 
dominance ordering on partitions of equal weight ($|\lambda|=|\mu|$): 
$\tablo{\lambda}\leq \tablo{\mu}
\Leftrightarrow \sum_i=1^j \lambda_i \leq \sum_{i=1}^j \mu_j$
for all $j$. The function $\jip{\nu}{\beta}$ in the inner product can
be shown to be given by \cite{McD}
\begin{eqnarray}
  \jip{\nu}{\beta}&=& \prod_{(i,j) \in \{\nu\}}
  {\beta(\nu'_j-i)+\nu_i-j+1\over \beta(\nu'_j-i+1)+\nu_i-j} \ .
\label{jlambda}
\end{eqnarray}

In a notation where Jack polynomials are written as functions
of power sums $p_n$, they satisfy a duality between
$\beta=p$ and $\beta={1\over p}$
\begin{equation}
  \Pjack{\lambda'}{p}{\left\{{p_n \over p}\right\}}=(-1)^{|\lambda|}{
    \jip{\lambda}{p}  }\Pjack{\lambda}{1\over p}{- \{p_n\}} \ ,
\label{jackdual}
\end{equation}
where $\tablo{\lambda'}$ is the Young tableau dual to $\tablo{\lambda}$.
It follows that
\begin{equation}
  \jip{\nu}{p} \, \jip{\nu'}{{1\over p}}=1 \ .
\end{equation}

The following elementary property of the Jack polynomials
\begin{equation}
  \label{eq:1-xyexpa}
  \prod_{i,j} (1-x_iy_j) =\sum_{\tablo{\lambda}} (-1)^{|\lambda|}
  \Pjack{\lambda}{\beta}{\{x_i\}}
  \Pjack{\lambda'}{{1\over \beta}}{\{y_j\}}
\end{equation}
can be used to rewrite expressions involving vertex-operators.

For integer $\beta$ an alternative inner product \cite{McD} on the
Jack polynomials $\PEjack{\mu}{\beta}$ 
depending on only a finite set of coordinates $\{z_i\}=\{z_1,...,z_n\}$
is given by
\begin{eqnarray}
\lefteqn{ \langle\langle\Pjack{\nu}{\beta}{\{z_i\}}|
   \Pjack{\mu}{\beta}{\{z_i\}} \rangle\rangle =}
\nonu
&& \left( \prod_{i=1}^{n}\oint{dz_i \over 2\pi i}{1 \over z_i} \right)
  \Delta^{\beta} ({\{{ z_i^{-1}}\}})
  \Delta^{\beta}({\{z_i\}})
  \Pjack{\nu}{\beta}{\{{ z_i^{-1}}\}}
  \Pjack{\mu}{\beta}{\{z_i\}} \ ,
  \label{eq:altip}
\end{eqnarray}
where $\Delta^\beta(\{x_i\})=\prod_{i<j} (x_i-x_j)^{\beta}$ denotes a
generalized Vandermonde determinant. Although it is also possible to
define this inner product for fractional $\beta$ \cite{McD}, we will
use it in this form for integer $\beta$. The Jack polynomials are
orthogonal w.r.t.\ this alternative inner product.

For the product of $N$ quasi-hole vertex operators
$\phi(z_i)$, the following expression can be derived
\begin{eqnarray}
  \lefteqn{\phi(z_1)...\phi(z_N)|q\rangle=}
\nonu
  &&
  \Delta^{1 \over p}(\{z_i\})
  \sum_{\tablo{\lambda}} (-1)^{|\lambda|} \,
  \Pjack{\lambda'}{{1\over p}}{\{z_j\}} \,
  \Jjack{\lambda}{p}{\left\{{a_{-n}\over \sqrt{p} }\right\}}
  \prod_{j=1}^N z_j^{q\over p} |q+N\rangle, 
\label{phiexpansion}
\end{eqnarray}
where we wrote
$\Jjack{\lambda}{p}{\left\{{a_{-n}\over \sqrt{p}}\right\}}$
for a Jack polynomials in which power sums $p_n$ are replaced by
bosonic modes, writing $a_n=(\del \varphi)_n$. [We refer to
such expressions as Jack operators.]
Similarly,
\begin{eqnarray}
  \lefteqn{J(z_1)...J(z_N)|q \rangle=}
\nonu
  && 
  \Delta^p(\{z_i\})
  \sum_{\tablo{\lambda}} (-1)^{|\lambda|}
  \Pjack{\lambda'}{p}{\{z_j\}}
  \Jjack{\lambda}{{1\over p}}{\{{\sqrt{p} a_{-n}}\}}
  \prod_{j=1}^N z_j^{-q} |q-Np\rangle   .
\label{Jexpansion}
\end{eqnarray}
For brevity, we sometimes drop the explicit reference to the bosonic
modes and write
\begin{equation}
  \JEjack{\lambda}{{1\over p}} \equiv
  \Jjack{\lambda}{{1\over p}}{\{{\sqrt{p} a_{-n}}\}}\nonu \ ,
\qquad
  \JEjack{\lambda}{{ p}} \equiv
  \Jjack{\lambda}{{ p}}{\{{{1\over \sqrt{p}} a_{-n}}\}}.
\end{equation}

\section{Proof of selection rules}
\setcounter{equation}{0}
\setcounter{figure}{0}

We present a proof of the form factor selection rules of section 4.3.

(a.) This is a consequence of energy conservation and can be found
from the product of three delta functions,
\begin{equation}
\delta_{|n|+m,\lambda_1}
\delta_{\lambda_1+|\rho|,|\mu|}\delta_{|n|+|\rho|,|\nu|}
\end{equation}
present implicitly in eq.~(\ref{eq:theorem}).

(b.)  Proposition 2.4 in~\cite{St} states that  it is possible to rewrite any
Jack polynomial as a linear combination of  products of Jack polynomials
labeled with horizontal
strips $\REpol{\lambda}^{p}=\prod_i P^p_{(\lambda_i)}$,
\begin{equation}
\PEjack{\lambda}{p}=
\sum_{\tablo{\sigma}\geq\tablo{\lambda}}\tilde{q}_{\tablo{\lambda}
  \tablo{\sigma}}
  \REpol{\sigma}^{p}.
\end{equation}
An important difference between this expansion and the expansion of a
Jack polynomial in monomial symmetric functions is that the sum runs
over tableaus $\tablo{\sigma}$ satisfying
$\tablo{\sigma}\geq\tablo{\lambda}$ instead of
$\tablo{\sigma}\leq\tablo{\lambda}$.  From repeated application of
proposition~5.3 in \cite{St} (see also section 4.3) 
to a product of two Jack polynomials, 
where one is expanded using the expansion above, it follows that the tableau
labeling the non-expanded Jack polynomial is contained in every
tableau labeling a Jack polynomial appearing in the product.
Exchanging the roles we see that also the tableau labeling the other
Jack polynomial is contained in these tableaus.

Combining this
knowledge with the fact that by triangularity the monomial symmetric
functions $m_{\tablo{n}}$ can be expanded in Jack polynomials,
\begin{equation}
  m_{\tablo{n}}= \sum_{\tablo{\tau}\leq\tablo{n}}
  \tilde{v}_{\tablo{n}\tablo{\tau}} \PEjack{\tau}{p},
\end{equation}  
and applying this to the coordinate inner product in
eq.~(\ref{eq:theorem}) we find that $\tablo{\rho}$ is 
contained in $\tablo{\nu}$. The operator inner product shows that 
$\tablo{\mu}$ differs at most a horizontal
strip from $\tablo{\rho}$ and thus $\tablo{\rho}$ contains
$\tablo{\tilde{\mu}}=(\mu_2,\dots,\mu_M)$. 
We can conclude that $\tablo{\nu}$ contains 
$\tablo{\tilde{\mu}}$ and we have obtained (b.1.).

We can extract
extra information from examining this construction once more, under
the addition of a horizontal strip the length of a column can grow
with one box. If we now multiply the Jack polynomials  $\PEjack{tau}{p}$
appearing in the expansion of the monomial symmetric function 
$m_{\tablo{n}}$ with the $\REpol{\sigma}^{p}$
appearing in the expansion of the Jack polynomial $\PEjack{\rho}{p}$
we find that the maximal difference in column length between
$\tablo{\nu}$ and $\tablo{\tau}$ is $l(\tablo{\rho})$ from which we
can conclude that the only columns in $\tablo{\nu}$ which have a length
exceeding $l(\tablo{\mu}) \geq l(\tablo{\rho}) $ are those
columns for which the corresponding tableau labeling the monomial
symmetric function has a column of  non-zero length. Because only monomial
symmetric functions with at most $p$ non-zero columns appear we obtain
(b.2.).

(c.) This is a simple consequence of (a.) and (b.1.)

(d.) By definition the Jack polynomials can
be expanded in monomial symmetric
functions, so we have  
\begin{equation}
  \PEjack{\rho}{p} =
  \sum_{\tablo{\sigma}\leq\tablo{\rho}} v_{\tablo{\rho}\tablo{\sigma}}
  m_{\tablo{\sigma}}.
\end{equation} 
If we now use this expansion $\PEjack{\rho}{p}$ and then multiply the
resulting $m_{\tablo{\sigma}}$ in the expansion by $m_{\tablo{n}}$,
then the products in the expansion will be linear combinations of
$m_{\tablo{\sigma'}}$ satisfying $\tablo{\sigma'}\leq
\tablo{\sigma}+\tablo{n}$,
\begin{eqnarray}
  m_{\tablo{n}}\PEjack{\rho}{p} &=&
  \sum_{\tablo{\sigma}\leq \tablo{\rho}} v_{\tablo{\rho}\tablo{\sigma}}
  m_{\tablo{n}}m_{\tablo{\sigma}} \nonu
  &=& \sum_{\tablo{\sigma'}\leq \tablo{\rho}+\tablo{n}}
  u_{\tablo{n}\tablo{\rho}\tablo{\sigma'}} m_{\tablo{\sigma'}}.
\end{eqnarray}

Expanding the $m_{\tablo{\sigma'}}$ as we
did in the proof of (b.) the sum over products of monomial symmetric
functions can be rewritten in terms of Jack polynomials again,
\begin{equation}
  m_{\tablo{n}}\PEjack{\rho}{p}=
  \sum_{\tablo{\sigma}\leq \tablo{\rho}+\tablo{n}}
  w_{\tablo{n}\tablo{\rho}\tablo{\sigma}} \PEjack{\sigma}{p} \ ,
\end{equation}
and we find that $\tablo{\nu}$ is smaller in the sense of dominance
order than a tableau $\tablo{\sigma}$ of the form
$\tablo{\rho}+\tablo{n}$. Since $n_i\leq p$ and
$\rho_i \leq \mu_i$, the result (d.) follows.

\newpage

\begin{center}
\setlength{\unitlength}{1mm}
\begin{picture}(140,140)(0,0)
\put(30,0)
{\epsfxsize=120mm{\epsffile{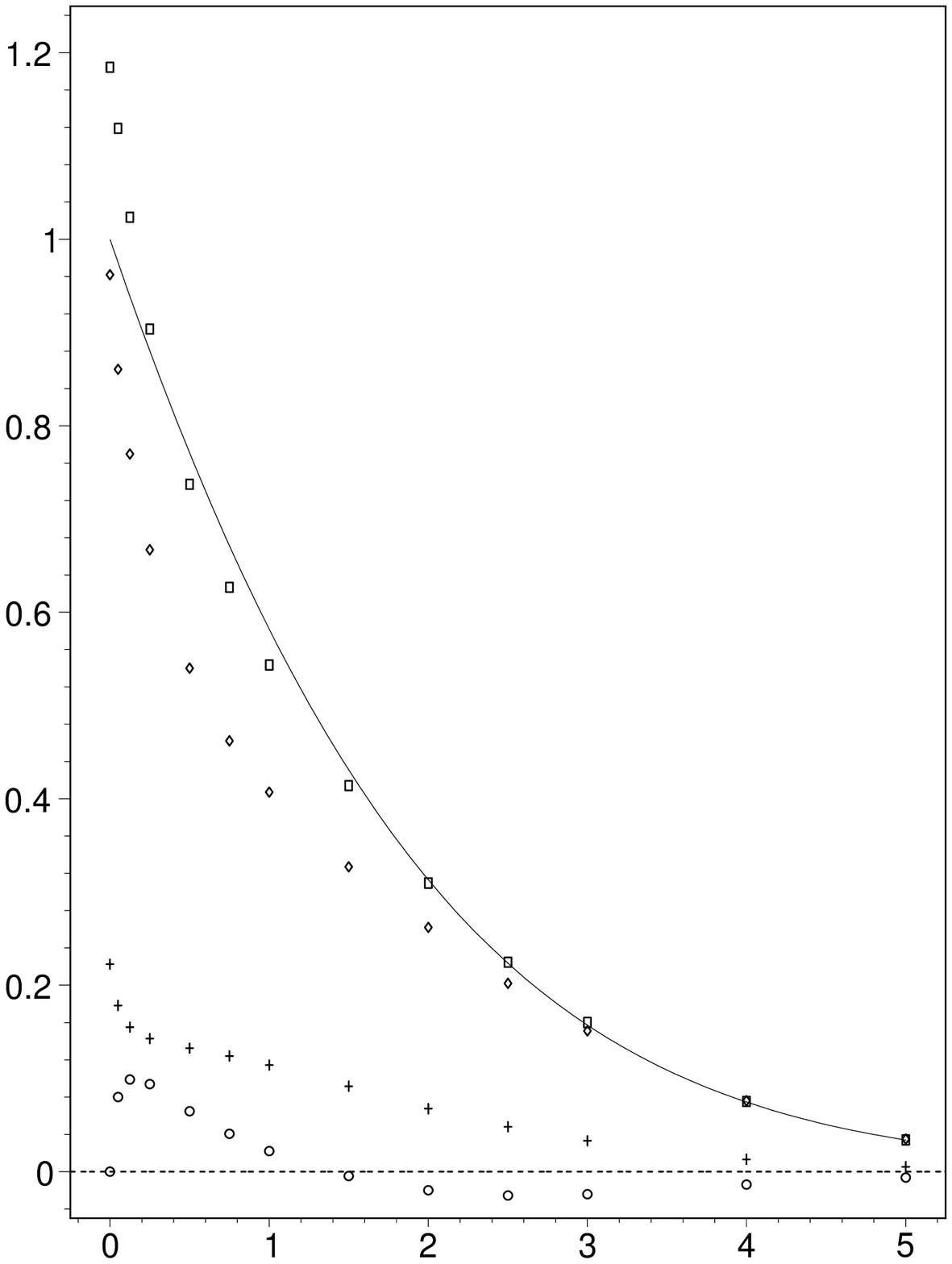}}}
\put(15,110){$h(\epsilon)$}
\put(90,0){$\epsilon \ [k_BT]$}
\end{picture}

\vskip 8mm

{\parbox{12.5cm}
{\small
   Figure 5.1:
   One-particle Green's function $h(\eps)$ for $p=2$
   as a function of energy. The solid curve is the exact result;
   the data points are the numerical results
   for: $h^{(1,0)}$ (diamonds), $h^{(2,0)}$ (circles) and
   $h^{(1,1)}$ (plusses). The sum of all contributions with
   up to two particles is represented by squares.}}
\end{center}

\end{document}